\documentclass[useAMS]{mn2e}
\usepackage{lscape}
\usepackage{rotating}

\def\one{\,{\sc i}}             
\def\two{\,{\sc ii}}
\def\three{\,{\sc iii}}

\input psfig.sty
\title[Ionizing Fluxes for Young Stellar Populations]
{Realistic Ionizing Fluxes for Young Stellar Populations from
0.05 to $\bmath{\bld{2} \times Z_\odot}$}
\author[L. J. Smith et al.] {Linda J. Smith\thanks{E-mail: 
ljs@star.ucl.ac.uk (LJS); rpfn@star.ucl.ac.uk (RPFN); pac@star.ucl.ac.uk (PAC)}, Richard P.F. Norris$^\star$ and Paul A. Crowther$^\star$\\
Department of Physics and
Astronomy, University College London, Gower Street, London, WC1E 6BT}
\date{Accepted. Received; in original form}

\pagerange{\pageref{firstpage}--\pageref{lastpage}}
\pubyear{2002}
\begin{document}

\maketitle

\label{firstpage}

\begin{abstract}
We present a new grid of ionizing fluxes for O and Wolf-Rayet stars
for use with evolutionary synthesis codes and single star H\two\
region analyses. A total of 230 expanding, non-LTE, line-blanketed model
atmospheres have been calculated for five metallicities (0.05, 0.2,
0.4, 1 and 2\,Z$_\odot$) using the WM-basic code of Pauldrach et al.
(2001) for O stars and the {\sc cmfgen} code of Hillier \& Miller
(1998) for W-R stars. The stellar wind parameters are scaled with
metallicity for both O and W-R stars. We compare the ionizing fluxes of
the new models with the CoStar models of Schaerer \& de Koter (1997)
and the pure helium W-R models of Schmutz, Leitherer \& Gruenwald
(1992).  We find significant differences, particularly above 54\,eV,
where the emergent flux is determined by the wind density as a
function of metallicity.  The new models have lower ionizing fluxes in
the He\one\ continuum with important implications for nebular line
ratios.  

We incorporate the new models into the evolutionary synthesis code
Starburst99 (Leitherer et al. 1999) and compare the ionizing outputs
for an instantaneous burst and continuous star formation with the work
of Schaerer \& Vacca (1998; SV98) and Leitherer et al. (1999).  The
changes in the output ionizing fluxes as a function of age are
dramatic. We find that, in contrast to previous studies, nebular
He\two\ $\lambda4686$ will be at, or just below, the detection limit in
low metallicity starbursts during the W-R phase. The new models have
lower fluxes in the He\one\ continuum for $Z \ge 0.4$\,Z$_\odot$ and
ages $\le 7$\,Myr because of the increased line blanketing.

We test the accuracy of the new model atmosphere grid by constructing
photoionization models for simple H\two\ regions, and assessing the
impact of the new ionizing fluxes on important nebular diagnostic line
ratios.  For the case of an H\two\ region where the ionizing flux is
given by the WM-basic dwarf O star grid, we show that He\one\
$\lambda5786$/H$\beta$ decreases between 1 and 2\,Z$_\odot$ in a
similar manner to observations (e.g. Bresolin et al. 1999).  We find
that this decline is caused by the increased effect of line blanketing
above solar metallicity. We therefore suggest that a lowering of the
upper mass limit at high abundances is not required to explain the
diminishing strength of He\one\ $\lambda5786$/H$\beta$, as has been
suggested in the past (e.g. Shields \& Tinsley 1976; Bresolin et
al. 1999). For an H\two\ region where the ionizing flux is provided by
an instantaneous burst of total mass $10^6$\,M$_\odot$, we plot the
softness parameter $\eta^\prime$ against the abundance indicator
$R_{23}$ for ages of 1--5\,Myr. The new models are coincident with the
observational data of Bresolin et al. (1999), particularly during the
W-R phase, unlike the previous models of SV98 which generally
over-predict the hardness of the ionizing radiation.

The new model grid and updated Starburst99 code can be downloaded from
http://www.star.ucl.ac.uk/starburst.
\end{abstract}

\section{Introduction}
Evolutionary synthesis codes are commonly used to derive the
properties of young, unresolved stellar populations from observations
at various wavelengths.  In the satellite ultraviolet, the stellar
wind spectral features of a population of massive stars can be
synthesized to provide information on the star formation rates, the
slope of the initial mass function (IMF), and ages (e.g. Robert,
Leitherer \& Heckman 1993; Leitherer, Robert \& Heckman 1995; de
Mello, Leitherer \& Heckman 2000). In the optical region, the
properties of integrated stellar populations are usually derived
indirectly from their total radiative energy outputs using nebular
diagnostic line ratios (e.g. Garc\'\i a-Vargas, Bressan \& D\'\i az
1995; Stasi\' nska \& Leitherer 1996; Stasi\' nska, Schaerer \&
Leitherer 2001).  This method uses the theoretical ionizing fluxes
from a population of massive stars as a function of age as input into
a photoionization code.  The ability of evolutionary synthesis models
to predict correctly the properties of a young stellar population from
nebular emission line ratios therefore depends heavily  on the accuracy
of the evolutionary and atmospheric models developed for single
massive stars.
 
Early attempts to model stellar populations using evolutionary
synthesis coupled with photoionization codes relied mainly on Kurucz (1992)
plane-parallel LTE model atmospheres (e.g.  Garc\'\i a-Vargas et
al. 1995). Gabler et al. (1989), however, showed that the
presence of a stellar wind has a significant effect on the emergent
ionizing flux in the neutral and ionized helium continua of O stars.
Non-LTE effects depopulate the ground state of He\two, leading
to a decrease in the bound-free opacity above 54\,eV, and hence a
larger flux in the He\two\ continuum by up to 3--6 orders of
magnitude. The emergent spectrum is flattened in the region of the
He\one\ continuum and has a higher flux due to the presence of the
stellar wind.

Stasi\' nska \& Leitherer (1996) were the first to use expanding
non-LTE atmospheres for stars with strong winds to construct
photoionization models for evolving starbursts. They used the grid of
pure helium, unblanketed non-LTE W-R atmospheres from Schmutz,
Leitherer \& Gruenwald (1992) to represent evolved stars with strong
winds, and Kurucz (1992) models for hot stars close to the main sequence.
Line-blanketed, expanding non-LTE models covering the main sequence evolution of
O stars were first introduced by Stasi\' nska \& Schaerer (1997) who
studied the effect of using the CoStar models of Schaerer \& de
Koter (1997) on the ionization of single star H\two\ regions. They found
that higher ionic ratios were obtained in comparison to Kurucz models
with the same stellar temperatures. Schaerer \& Vacca (1998) used the
CoStar and Schmutz et al. (1992) grids to construct evolutionary
synthesis models for young starbursts. They predicted strong nebular
He\two\ $\lambda4686$ in low metallicity starbursts containing W-R
stars. The same evolutionary synthesis models were used by Stasi\'
nska, Schaerer \& Leitherer (2001) combined with photoionization models
to analyse the emission line properties of H\two\ galaxies.

One major question arises: how realistic are the ionizing fluxes being
used in the evolutionary synthesis studies outlined above? To address
this, there have been numerous studies aimed at empirically testing
the accuracy of hot star model atmospheres by analysing H\two\ regions
containing single stars with well defined spectral types and effective
temperatures.  Schaerer (2000) reviews the success of non-LTE model
atmospheres in reproducing nebular diagnostic line ratios.  Esteban et
al. (1993) examined the accuracy of the W-R grids by photoionization
modelling of nebulae ionized by single W-R stars. They found that the
lack of line blanketing was most important for the coolest W-R
stars. In recent years, new computational techniques have allowed
line-blanketed W-R atmospheres to be calculated for a few stars
(e.g. Schmutz 1997; de Koter, Heap \& Hubeny 1997; Hillier \& Miller
1998).  Crowther et al. (1999) tested these models by seeking to find
a consistent model that reproduced the stellar and nebular parameters
of a cool WN star.  They found that line blanketing plays a
significant role in modifying the W-R ionizing output although the
models of de Koter et al. (1997) and Hillier \& Miller (1998) predict
quite different ionizing flux distributions below the He\one\ edge.

Bresolin, Kennicutt \& Garnett (1999) modelled extragalactic H\two\
regions using Kurucz and non-LTE models. They find that the
temperatures of the ionizing stars decrease with increasing
metallicity, and suggest that this can be explained by lowering the
upper mass limit for star formation. In a further study, Kennicutt et
al. (2000) have analysed a large sample of H\two\ regions containing
stars of known spectral types and effective temperatures.  They
confirm the dependence of stellar temperature on metallicity, although
they note that this relationship depends on the correctness of the
input ionizing fluxes, and particularly the amount of line blanketing.
Oey et al. (2000) have presented a detailed comparison of spatially
resolved H\two\ region spectra to photoionization models to test how
well the CoStar models and Schmutz et al. (1992) W-R models reproduce
the observed nebular line ratios. They find that overall the agreement
is within 0.2\,dex but the nebular models appear to be too hot by
$\sim 1000$\,K, and suggest that the ionizing flux distribution may be
too hard in the 41--54\,eV range.  On the other hand, Kewley et
al. (2001) model a large sample of infrared starburst galaxies, and
find that the Schmutz et al. (1992) W-R atmospheres do not produce
sufficient flux between 13.6--54\,eV to match their observations.

It is clear that the use of expanding non-LTE model atmospheres in the
analysis of single and unresolved H\two\ regions is a major
improvement over using static LTE models to represent hot massive
stars. The empirical studies described above show that, in addition,
it is essential that the atmospheres are line-blanketed, particularly
for studies of young stellar populations at solar or higher
metallicities.  The CoStar models used in more recent studies do
incorporate line blanketing but the lack of any line blanketing in the
W-R atmosphere grid of Schmutz et al. (1992) is a serious deficiency.
With the recent advances in computing and the development of large
codes to calculate expanding, non-LTE line blanketed atmospheres, it is now
feasible to compute a grid of realistic ionizing fluxes for OB and W-R
stars.  In this paper, we present such a grid for five metallicities
from 0.05--2\,Z$_\odot$ for use with the Starburst99 (Leitherer et
al. 1999) evolutionary synthesis code and analyses of single star
H\two\ regions.  In Section~\ref{grid}, we present the details of the
model atmosphere grid, and in Section~\ref{comp}, we compare the
predicted ionizing fluxes with previous single star models used in
synthesis codes. In Section~\ref{sb99}, we incorporate the new grid
into Starburst99, and in Section~\ref{evolcomp}, we examine the
changes in the ionizing fluxes as a function of age for an
instantaneous burst and continuous star formation. In
Section~\ref{photo}, we use the photoionization code {\sc cloudy}
(Ferland 2002) to assess the differences in the nebular diagnostic
line ratios of H\two\ regions ionized by single O stars and synthetic
clusters as a function of age. We discuss our results in
Section~\ref{discuss} and present the conclusions in Section~\ref{conc}.

\section{The Model Atmosphere Grid}\label{grid}
We have sought to compute a grid of expanding, non-LTE, line-blanketed model
atmospheres that cover the entire upper H-R diagram for massive stars
with stellar winds.  To this end, we have used two model atmosphere
codes: the WM-basic code of Pauldrach et al. (2001) for O and early B
stars; and the {\sc cmfgen} code of Hillier and Miller (1998) for W-R
stars.  In total, we have calculated 230 model atmospheres using a 350
MHz Pentium II PC for WM-basic and a 1.3 GHz Pentium IV PC for {\sc
cmfgen}. Each O and W-R model calculation typically took 3 and 9 hours
to complete respectively.  We have calculated the models for five
metallicities: 0.05, 0.2, 0.4, 1, and 2\,Z$_\odot$, as defined by the
evolutionary tracks of Meynet et al. (1994).
\begin{sidewaystable*}
\label{tab_OBV}
\begin{tabular}{@{}*{4}{c}l*{6}{c}*{6}{c}*{6}{c}*{6}{c}*{6}{c}}
\noalign{{\bf Table 1:} OB Dwarf Model Atmosphere Grid}\\
\hline
&&&&&
\multicolumn{6}{c}{ \bf{Z$_{\odot}$}} &
\multicolumn{6}{c}{\bf{0.2\,Z$_{\odot}$}} &
\multicolumn{6}{c}{\bf{0.4\,Z$_{\odot}$}} &
\multicolumn{6}{c}{\bf{0.05\,Z$_{\odot}$}} &
\multicolumn{6}{c}{\bf{2.0\,Z$_{\odot}$}} \\ 
Model & $T_{\rm {eff}}$ & $\log g$ & $R_{\star}$ & SpT
& \multicolumn{3}{c}{log
$\dot{M}$}&\multicolumn{3}{c}{$v_{\infty}$}&\multicolumn{3}{c}{log
$\dot{M}$}&\multicolumn{3}{c}{$v_{\infty}$}&\multicolumn{3}{c}{log
$\dot{M}$}&\multicolumn{3}{c}{$v_{\infty}$}&\multicolumn{3}{c}{log
$\dot{M}$}&\multicolumn{3}{c}{$v_{\infty}$}&\multicolumn{3}{c}{log
$\dot{M}$}&\multicolumn{3}{c}{$v_{\infty}$}\\
Ref. &(kK) & (cm\,s$^{-2}$)& (R$_{\odot}$)&&
\multicolumn{2}{c}{ Q$_{0}$}&\multicolumn{2}{c}{
  Q$_{1}$}&\multicolumn{2}{c}{ Q$_{2}$}&\multicolumn{2}{c}{
  Q$_{0}$}&\multicolumn{2}{c}{ Q$_{1}$}&\multicolumn{2}{c}{
  Q$_{2}$}&\multicolumn{2}{c}{ Q$_{0}$}&\multicolumn{2}{c}{
  Q$_{1}$}&\multicolumn{2}{c}{ Q$_{2}$}&\multicolumn{2}{c}{
  Q$_{0}$}&\multicolumn{2}{c}{ Q$_{1}$}&\multicolumn{2}{c}{
  Q$_{2}$}&\multicolumn{2}{c}{ Q$_{0}$}&\multicolumn{2}{c}{
  Q$_{1}$}&\multicolumn{2}{c}{ Q$_{2}$}\\
\hline
 OB\#1&   50.0&4.00& 9.8& O3V&\multicolumn{3}{c}{  -5.85}&\multicolumn{3}{c}{3150}&\multicolumn{3}{c}{  -6.41}&\multicolumn{3}{c}{2550}&\multicolumn{3}{c}{  -6.17}&\multicolumn{3}{c}{2790}&\multicolumn{3}{c}{  -6.89}&\multicolumn{3}{c}{2130}&\multicolumn{3}{c}{  -5.61}&\multicolumn{3}{c}{3440} \\

&&&&&\multicolumn{2}{c}{49.5}&\multicolumn{2}{c}{48.8}&\multicolumn{2}{c}{45.8}&\multicolumn{2}{c}{49.5}&\multicolumn{2}{c}{49.0}&\multicolumn{2}{c}{45.6}&\multicolumn{2}{c}{49.5}&\multicolumn{2}{c}{48.9}&\multicolumn{2}{c}{45.7}&\multicolumn{2}{c}{49.6}&\multicolumn{2}{c}{49.1}&\multicolumn{2}{c}{45.4}&\multicolumn{2}{c}{49.5}&\multicolumn{2}{c}{48.8}&\multicolumn{2}{c}{45.6} \\

 OB\#2&   45.7&4.00&10.4& O4V &\multicolumn{3}{c}{  -6.08}&\multicolumn{3}{c}{2950}&\multicolumn{3}{c}{  -6.63}&\multicolumn{3}{c}{2390}&\multicolumn{3}{c}{  -6.39}&\multicolumn{3}{c}{2610}&\multicolumn{3}{c}{  -7.12}&\multicolumn{3}{c}{1990}&\multicolumn{3}{c}{  -5.83}&\multicolumn{3}{c}{3220} \\

&&&&&\multicolumn{2}{c}{49.4}&\multicolumn{2}{c}{48.7}&\multicolumn{2}{c}{45.0}&\multicolumn{2}{c}{49.4}&\multicolumn{2}{c}{48.7}&\multicolumn{2}{c}{45.2}&\multicolumn{2}{c}{49.4}&\multicolumn{2}{c}{48.7}&\multicolumn{2}{c}{45.2}&\multicolumn{2}{c}{49.4}&\multicolumn{2}{c}{48.8}&\multicolumn{2}{c}{45.0}&\multicolumn{2}{c}{49.4}&\multicolumn{2}{c}{48.6}&\multicolumn{2}{c}{44.8} \\

 OB\#3&   42.6&4.00&10.5& O5V &\multicolumn{3}{c}{  -6.24}&\multicolumn{3}{c}{2870}&\multicolumn{3}{c}{  -6.80}&\multicolumn{3}{c}{2330}&\multicolumn{3}{c}{  -6.55}&\multicolumn{3}{c}{2550}&\multicolumn{3}{c}{  -7.28}&\multicolumn{3}{c}{1940}&\multicolumn{3}{c}{  -6.00}&\multicolumn{3}{c}{3140} \\

&&&&&\multicolumn{2}{c}{49.2}&\multicolumn{2}{c}{48.5}&\multicolumn{2}{c}{44.4}&\multicolumn{2}{c}{49.2}&\multicolumn{2}{c}{48.6}&\multicolumn{2}{c}{44.2}&\multicolumn{2}{c}{49.2}&\multicolumn{2}{c}{48.5}&\multicolumn{2}{c}{44.3}&\multicolumn{2}{c}{49.2}&\multicolumn{2}{c}{48.6}&\multicolumn{2}{c}{44.2}&\multicolumn{2}{c}{49.2}&\multicolumn{2}{c}{48.5}&\multicolumn{2}{c}{44.5} \\

 OB\#4&   40.0&4.00&10.6&  O7V&\multicolumn{3}{c}{  -6.34}&\multicolumn{3}{c}{2500}&\multicolumn{3}{c}{  -6.90}&\multicolumn{3}{c}{2020}&\multicolumn{3}{c}{  -6.66}&\multicolumn{3}{c}{2210}&\multicolumn{3}{c}{  -7.38}&\multicolumn{3}{c}{1690}&\multicolumn{3}{c}{  -6.10}&\multicolumn{3}{c}{2730} \\

&&&&&\multicolumn{2}{c}{49.0}&\multicolumn{2}{c}{48.2}&\multicolumn{2}{c}{43.8}&\multicolumn{2}{c}{49.0}&\multicolumn{2}{c}{48.3}&\multicolumn{2}{c}{43.4}&\multicolumn{2}{c}{49.0}&\multicolumn{2}{c}{48.3}&\multicolumn{2}{c}{43.5}&\multicolumn{2}{c}{49.0}&\multicolumn{2}{c}{48.4}&\multicolumn{2}{c}{43.6}&\multicolumn{2}{c}{49.0}&\multicolumn{2}{c}{48.1}&\multicolumn{2}{c}{37.7} \\

 OB\#5&   37.2&4.00&10.5& O7.5V&\multicolumn{3}{c}{  -6.47}&\multicolumn{3}{c}{2100}&\multicolumn{3}{c}{  -7.03}&\multicolumn{3}{c}{1700}&\multicolumn{3}{c}{  -6.79}&\multicolumn{3}{c}{1860}&\multicolumn{3}{c}{  -7.51}&\multicolumn{3}{c}{1420}&\multicolumn{3}{c}{  -6.23}&\multicolumn{3}{c}{2290} \\

&&&&&\multicolumn{2}{c}{48.7}&\multicolumn{2}{c}{47.6}&\multicolumn{2}{c}{37.3}&\multicolumn{2}{c}{48.8}&\multicolumn{2}{c}{47.8}&\multicolumn{2}{c}{42.3}&\multicolumn{2}{c}{48.8}&\multicolumn{2}{c}{47.8}&\multicolumn{2}{c}{42.4}&\multicolumn{2}{c}{48.7}&\multicolumn{2}{c}{47.8}&\multicolumn{2}{c}{42.5}&\multicolumn{2}{c}{48.7}&\multicolumn{2}{c}{47.4}&\multicolumn{2}{c}{38.5} \\

 OB\#6&   34.6&4.00&10.5&  O8V&\multicolumn{3}{c}{  -6.64}&\multicolumn{3}{c}{1950}&\multicolumn{3}{c}{  -7.20}&\multicolumn{3}{c}{1580}&\multicolumn{3}{c}{  -6.96}&\multicolumn{3}{c}{1730}&\multicolumn{3}{c}{  -7.68}&\multicolumn{3}{c}{1320}&\multicolumn{3}{c}{  -6.40}&\multicolumn{3}{c}{2130} \\

&&&&&\multicolumn{2}{c}{48.5}&\multicolumn{2}{c}{46.6}&\multicolumn{2}{c}{36.3}&\multicolumn{2}{c}{48.4}&\multicolumn{2}{c}{46.9}&\multicolumn{2}{c}{36.7}&\multicolumn{2}{c}{48.4}&\multicolumn{2}{c}{46.8}&\multicolumn{2}{c}{36.7}&\multicolumn{2}{c}{48.4}&\multicolumn{2}{c}{46.7}&\multicolumn{2}{c}{38.7}&\multicolumn{2}{c}{48.4}&\multicolumn{2}{c}{46.7}&\multicolumn{2}{c}{35.4} \\

 OB\#7&   32.3&4.00&10.1& O9V&\multicolumn{3}{c}{  -6.74}&\multicolumn{3}{c}{1500}&\multicolumn{3}{c}{  -7.30}&\multicolumn{3}{c}{1210}&\multicolumn{3}{c}{  -7.06}&\multicolumn{3}{c}{1330}&\multicolumn{3}{c}{  -7.79}&\multicolumn{3}{c}{1010}&\multicolumn{3}{c}{  -6.50}&\multicolumn{3}{c}{1640} \\

&&&&&\multicolumn{2}{c}{47.9}&\multicolumn{2}{c}{45.7}&\multicolumn{2}{c}{34.3}&\multicolumn{2}{c}{48.0}&\multicolumn{2}{c}{45.5}&\multicolumn{2}{c}{34.8}&\multicolumn{2}{c}{47.9}&\multicolumn{2}{c}{45.5}&\multicolumn{2}{c}{34.9}&\multicolumn{2}{c}{48.0}&\multicolumn{2}{c}{45.6}&\multicolumn{2}{c}{34.8}&\multicolumn{2}{c}{48.0}&\multicolumn{2}{c}{45.6}&\multicolumn{2}{c}{33.6} \\

 OB\#8&   30.2&4.00& 9.5& B0V &\multicolumn{3}{c}{  -6.92}&\multicolumn{3}{c}{1200}&\multicolumn{3}{c}{  -7.48}&\multicolumn{3}{c}{ 970}&\multicolumn{3}{c}{  -7.24}&\multicolumn{3}{c}{1060}&\multicolumn{3}{c}{  -7.96}&\multicolumn{3}{c}{ 810}&\multicolumn{3}{c}{  -6.68}&\multicolumn{3}{c}{1310} \\

&&&&&\multicolumn{2}{c}{47.4}&\multicolumn{2}{c}{44.8}&\multicolumn{2}{c}{32.7}&\multicolumn{2}{c}{47.4}&\multicolumn{2}{c}{44.6}&\multicolumn{2}{c}{33.3}&\multicolumn{2}{c}{47.3}&\multicolumn{2}{c}{44.8}&\multicolumn{2}{c}{33.4}&\multicolumn{2}{c}{47.2}&\multicolumn{2}{c}{44.5}&\multicolumn{2}{c}{33.2}&\multicolumn{2}{c}{47.4}&\multicolumn{2}{c}{44.9}&\multicolumn{2}{c}{32.7} \\

 OB\#9&   28.1&4.00& 8.6& B0.5V&\multicolumn{3}{c}{  -6.74}&\multicolumn{3}{c}{ 800}&\multicolumn{3}{c}{  -7.30}&\multicolumn{3}{c}{ 640}&\multicolumn{3}{c}{  -7.06}&\multicolumn{3}{c}{ 710}&\multicolumn{3}{c}{  -7.79}&\multicolumn{3}{c}{ 540}&\multicolumn{3}{c}{  -6.50}&\multicolumn{3}{c}{ 870} \\

&&&&&\multicolumn{2}{c}{47.0}&\multicolumn{2}{c}{44.5}&\multicolumn{2}{c}{31.5}&\multicolumn{2}{c}{46.8}&\multicolumn{2}{c}{44.2}&\multicolumn{2}{c}{32.4}&\multicolumn{2}{c}{46.8}&\multicolumn{2}{c}{44.3}&\multicolumn{2}{c}{32.2}&\multicolumn{2}{c}{46.8}&\multicolumn{2}{c}{43.9}&\multicolumn{2}{c}{32.1}&\multicolumn{2}{c}{46.9}&\multicolumn{2}{c}{44.5}&\multicolumn{2}{c}{31.1} \\

 OB\#10&   26.3&4.00& 8.1& B1V&\multicolumn{3}{c}{  -6.89}&\multicolumn{3}{c}{ 700}&\multicolumn{3}{c}{  -7.45}&\multicolumn{3}{c}{ 560}&\multicolumn{3}{c}{  -7.20}&\multicolumn{3}{c}{ 620}&\multicolumn{3}{c}{  -7.93}&\multicolumn{3}{c}{ 470}&\multicolumn{3}{c}{  -6.65}&\multicolumn{3}{c}{ 760} \\

&&&&&\multicolumn{2}{c}{46.5}&\multicolumn{2}{c}{43.9}&\multicolumn{2}{c}{30.5}&\multicolumn{2}{c}{46.3}&\multicolumn{2}{c}{43.7}&\multicolumn{2}{c}{31.6}&\multicolumn{2}{c}{46.4}&\multicolumn{2}{c}{43.8}&\multicolumn{2}{c}{31.4}&\multicolumn{2}{c}{46.2}&\multicolumn{2}{c}{43.3}&\multicolumn{2}{c}{30.9}&\multicolumn{2}{c}{46.4}&\multicolumn{2}{c}{44.1}&\multicolumn{2}{c}{...} \\
 OB\#11&   25.0&4.00& 7.1& B1.5V&\multicolumn{3}{c}{  -7.08}&\multicolumn{3}{c}{ 600}&\multicolumn{3}{c}{  -7.63}&\multicolumn{3}{c}{ 480}&\multicolumn{3}{c}{  -7.39}&\multicolumn{3}{c}{ 530}&\multicolumn{3}{c}{  -8.12}&\multicolumn{3}{c}{ 400}&\multicolumn{3}{c}{  -6.83}&\multicolumn{3}{c}{ 650} \\
&&&&&\multicolumn{2}{c}{46.1}&\multicolumn{2}{c}{43.4}&\multicolumn{2}{c}{...}&\multicolumn{2}{c}{45.9}&\multicolumn{2}{c}{43.2}&\multicolumn{2}{c}{30.8}&\multicolumn{2}{c}{46.1}&\multicolumn{2}{c}{43.3}&\multicolumn{2}{c}{30.9}&\multicolumn{2}{c}{45.8}&\multicolumn{2}{c}{42.9}&\multicolumn{2}{c}{...}&\multicolumn{2}{c}{46.1}&\multicolumn{2}{c}{43.7}&\multicolumn{2}{c}{...} \\
\hline\\
\noalign{
Notes: Mass loss rates $\dot{M}$ and terminal velocities $v_{\infty}$ are in units of
M$_{\odot}$\,yr$^{-1}$ and km\,s$^{-1}$ respectively. $Q_0, Q_1$ and $Q_2$ are on
a log scale and represent the photon luminosity (s$^{-1}$) in the
H\one, He\one\ and He\two\ ionizing continua.}
\end{tabular}
\end{sidewaystable*}
\begin{sidewaystable*}
\label{tab_OBI}
\begin{tabular}{@{}*{4}{c}l*{6}{c}*{6}{c}*{6}{c}*{6}{c}*{6}{c}}
\noalign{{\bf Table 2:} OB Supergiant Model Atmosphere Grid}\\
\hline
&&&&&
\multicolumn{6}{c}{ \bf{Z$_{\odot}$}} &
\multicolumn{6}{c}{\bf{0.2\,Z$_{\odot}$}} &
\multicolumn{6}{c}{\bf{0.4\,Z$_{\odot}$}} &
\multicolumn{6}{c}{\bf{0.05\,Z$_{\odot}$}} &
\multicolumn{6}{c}{\bf{2.0\,Z$_{\odot}$}} \\ 
Model & $T_{\rm {eff}}$ & $\log g$ & $R_{\star}$ & SpT
& \multicolumn{3}{c}{log
$\dot{M}$}&\multicolumn{3}{c}{$v_{\infty}$}&\multicolumn{3}{c}{log
$\dot{M}$}&\multicolumn{3}{c}{$v_{\infty}$}&\multicolumn{3}{c}{log
$\dot{M}$}&\multicolumn{3}{c}{$v_{\infty}$}&\multicolumn{3}{c}{log
$\dot{M}$}&\multicolumn{3}{c}{$v_{\infty}$}&\multicolumn{3}{c}{log
$\dot{M}$}&\multicolumn{3}{c}{$v_{\infty}$}\\
Ref. &(kK) & (cm\,s$^{-2}$)& (R$_{\odot}$)&&
\multicolumn{2}{c}{ Q$_{0}$}&\multicolumn{2}{c}{
  Q$_{1}$}&\multicolumn{2}{c}{ Q$_{2}$}&\multicolumn{2}{c}{
  Q$_{0}$}&\multicolumn{2}{c}{ Q$_{1}$}&\multicolumn{2}{c}{
  Q$_{2}$}&\multicolumn{2}{c}{ Q$_{0}$}&\multicolumn{2}{c}{
  Q$_{1}$}&\multicolumn{2}{c}{ Q$_{2}$}&\multicolumn{2}{c}{
  Q$_{0}$}&\multicolumn{2}{c}{ Q$_{1}$}&\multicolumn{2}{c}{
  Q$_{2}$}&\multicolumn{2}{c}{ Q$_{0}$}&\multicolumn{2}{c}{
  Q$_{1}$}&\multicolumn{2}{c}{ Q$_{2}$}\\
\hline
 OB\#12&   51.5&3.88&15.2& O3I&\multicolumn{3}{c}{  -4.59}&\multicolumn{3}{c}{3150}&\multicolumn{3}{c}{  -5.14}&\multicolumn{3}{c}{2550}&\multicolumn{3}{c}{  -4.90}&\multicolumn{3}{c}{2790}&\multicolumn{3}{c}{  -5.63}&\multicolumn{3}{c}{2130}&\multicolumn{3}{c}{  -4.34}&\multicolumn{3}{c}{3440} \\

&&&&&\multicolumn{2}{c}{50.0}&\multicolumn{2}{c}{49.3}&\multicolumn{2}{c}{39.4}&\multicolumn{2}{c}{50.0}&\multicolumn{2}{c}{49.5}&\multicolumn{2}{c}{46.7}&\multicolumn{2}{c}{50.0}&\multicolumn{2}{c}{49.4}&\multicolumn{2}{c}{46.6}&\multicolumn{2}{c}{50.0}&\multicolumn{2}{c}{49.5}&\multicolumn{2}{c}{46.4}&\multicolumn{2}{c}{50.0}&\multicolumn{2}{c}{49.2}&\multicolumn{2}{c}{39.9} \\

 OB\#13&   45.7&3.73&18.4& O4I&\multicolumn{3}{c}{  -4.72}&\multicolumn{3}{c}{2320}&\multicolumn{3}{c}{  -5.28}&\multicolumn{3}{c}{1880}&\multicolumn{3}{c}{  -5.04}&\multicolumn{3}{c}{2060}&\multicolumn{3}{c}{  -5.76}&\multicolumn{3}{c}{1570}&\multicolumn{3}{c}{  -4.48}&\multicolumn{3}{c}{2540} \\

&&&&&\multicolumn{2}{c}{49.9}&\multicolumn{2}{c}{49.0}&\multicolumn{2}{c}{38.2}&\multicolumn{2}{c}{49.9}&\multicolumn{2}{c}{49.2}&\multicolumn{2}{c}{45.6}&\multicolumn{2}{c}{49.9}&\multicolumn{2}{c}{49.2}&\multicolumn{2}{c}{38.9}&\multicolumn{2}{c}{49.9}&\multicolumn{2}{c}{49.3}&\multicolumn{2}{c}{45.9}&\multicolumn{2}{c}{50.0}&\multicolumn{2}{c}{48.9}&\multicolumn{2}{c}{37.2} \\

 OB\#14&   42.6&3.67&19.7& O5I&\multicolumn{3}{c}{  -4.92}&\multicolumn{3}{c}{2300}&\multicolumn{3}{c}{  -5.48}&\multicolumn{3}{c}{1860}&\multicolumn{3}{c}{  -5.24}&\multicolumn{3}{c}{2040}&\multicolumn{3}{c}{  -5.96}&\multicolumn{3}{c}{1550}&\multicolumn{3}{c}{  -4.68}&\multicolumn{3}{c}{2510} \\

&&&&&\multicolumn{2}{c}{49.8}&\multicolumn{2}{c}{48.9}&\multicolumn{2}{c}{38.3}&\multicolumn{2}{c}{49.8}&\multicolumn{2}{c}{49.1}&\multicolumn{2}{c}{45.4}&\multicolumn{2}{c}{49.9}&\multicolumn{2}{c}{49.1}&\multicolumn{2}{c}{45.2}&\multicolumn{2}{c}{49.8}&\multicolumn{2}{c}{49.2}&\multicolumn{2}{c}{45.4}&\multicolumn{2}{c}{49.8}&\multicolumn{2}{c}{48.7}&\multicolumn{2}{c}{37.6} \\

 OB\#15&   40.0&3.51&21.5& O6I&\multicolumn{3}{c}{  -5.01}&\multicolumn{3}{c}{2180}&\multicolumn{3}{c}{  -5.57}&\multicolumn{3}{c}{1760}&\multicolumn{3}{c}{  -5.33}&\multicolumn{3}{c}{1930}&\multicolumn{3}{c}{  -6.05}&\multicolumn{3}{c}{1470}&\multicolumn{3}{c}{  -4.77}&\multicolumn{3}{c}{2380} \\

&&&&&\multicolumn{2}{c}{49.8}&\multicolumn{2}{c}{48.8}&\multicolumn{2}{c}{38.0}&\multicolumn{2}{c}{49.8}&\multicolumn{2}{c}{49.0}&\multicolumn{2}{c}{44.5}&\multicolumn{2}{c}{49.8}&\multicolumn{2}{c}{49.0}&\multicolumn{2}{c}{38.1}&\multicolumn{2}{c}{49.8}&\multicolumn{2}{c}{49.1}&\multicolumn{2}{c}{38.2}&\multicolumn{2}{c}{49.7}&\multicolumn{2}{c}{48.6}&\multicolumn{2}{c}{36.9} \\

 OB\#16&   37.2&3.40&23.6& O7.5I&\multicolumn{3}{c}{  -5.07}&\multicolumn{3}{c}{1980}&\multicolumn{3}{c}{  -5.62}&\multicolumn{3}{c}{1600}&\multicolumn{3}{c}{  -5.38}&\multicolumn{3}{c}{1750}&\multicolumn{3}{c}{  -6.11}&\multicolumn{3}{c}{1340}&\multicolumn{3}{c}{  -4.82}&\multicolumn{3}{c}{2160} \\

&&&&&\multicolumn{2}{c}{49.6}&\multicolumn{2}{c}{48.5}&\multicolumn{2}{c}{37.2}&\multicolumn{2}{c}{49.7}&\multicolumn{2}{c}{48.9}&\multicolumn{2}{c}{37.6}&\multicolumn{2}{c}{49.7}&\multicolumn{2}{c}{48.8}&\multicolumn{2}{c}{37.7}&\multicolumn{2}{c}{49.7}&\multicolumn{2}{c}{48.9}&\multicolumn{2}{c}{44.4}&\multicolumn{2}{c}{49.6}&\multicolumn{2}{c}{48.3}&\multicolumn{2}{c}{36.0} \\

 OB\#17&   34.6&3.29&24.6& O8.5I&\multicolumn{3}{c}{  -5.19}&\multicolumn{3}{c}{1950}&\multicolumn{3}{c}{  -5.75}&\multicolumn{3}{c}{1580}&\multicolumn{3}{c}{  -5.51}&\multicolumn{3}{c}{1730}&\multicolumn{3}{c}{  -6.23}&\multicolumn{3}{c}{1320}&\multicolumn{3}{c}{  -4.95}&\multicolumn{3}{c}{2130} \\

&&&&&\multicolumn{2}{c}{49.5}&\multicolumn{2}{c}{48.3}&\multicolumn{2}{c}{36.3}&\multicolumn{2}{c}{49.5}&\multicolumn{2}{c}{48.5}&\multicolumn{2}{c}{37.3}&\multicolumn{2}{c}{49.5}&\multicolumn{2}{c}{48.5}&\multicolumn{2}{c}{37.3}&\multicolumn{2}{c}{49.5}&\multicolumn{2}{c}{48.6}&\multicolumn{2}{c}{37.4}&\multicolumn{2}{c}{49.4}&\multicolumn{2}{c}{47.9}&\multicolumn{2}{c}{35.0} \\

 OB\#18&   32.3&3.23&25.5& O9.5I&\multicolumn{3}{c}{  -5.36}&\multicolumn{3}{c}{1990}&\multicolumn{3}{c}{  -5.92}&\multicolumn{3}{c}{1610}&\multicolumn{3}{c}{  -5.67}&\multicolumn{3}{c}{1760}&\multicolumn{3}{c}{  -6.40}&\multicolumn{3}{c}{1340}&\multicolumn{3}{c}{  -5.12}&\multicolumn{3}{c}{2170} \\

&&&&&\multicolumn{2}{c}{49.3}&\multicolumn{2}{c}{47.5}&\multicolumn{2}{c}{33.9}&\multicolumn{2}{c}{49.3}&\multicolumn{2}{c}{47.8}&\multicolumn{2}{c}{36.7}&\multicolumn{2}{c}{49.3}&\multicolumn{2}{c}{47.8}&\multicolumn{2}{c}{36.2}&\multicolumn{2}{c}{49.3}&\multicolumn{2}{c}{48.0}&\multicolumn{2}{c}{36.8}&\multicolumn{2}{c}{49.2}&\multicolumn{2}{c}{47.3}&\multicolumn{2}{c}{33.7} \\

 OB\#19&   30.2&3.14&28.1& O9.7I&\multicolumn{3}{c}{  -5.37}&\multicolumn{3}{c}{1800}&\multicolumn{3}{c}{  -5.93}&\multicolumn{3}{c}{1460}&\multicolumn{3}{c}{  -5.68}&\multicolumn{3}{c}{1590}&\multicolumn{3}{c}{  -6.41}&\multicolumn{3}{c}{1210}&\multicolumn{3}{c}{  -5.13}&\multicolumn{3}{c}{1960} \\

&&&&&\multicolumn{2}{c}{49.1}&\multicolumn{2}{c}{46.4}&\multicolumn{2}{c}{32.8}&\multicolumn{2}{c}{49.0}&\multicolumn{2}{c}{46.6}&\multicolumn{2}{c}{34.6}&\multicolumn{2}{c}{49.0}&\multicolumn{2}{c}{46.4}&\multicolumn{2}{c}{34.0}&\multicolumn{2}{c}{49.1}&\multicolumn{2}{c}{47.1}&\multicolumn{2}{c}{35.5}&\multicolumn{2}{c}{49.0}&\multicolumn{2}{c}{46.4}&\multicolumn{2}{c}{...} \\

 OB\#20&   28.1&3.08&27.8& B0I&\multicolumn{3}{c}{  -5.89}&\multicolumn{3}{c}{1500}&\multicolumn{3}{c}{  -6.45}&\multicolumn{3}{c}{1210}&\multicolumn{3}{c}{  -6.20}&\multicolumn{3}{c}{1330}&\multicolumn{3}{c}{  -6.93}&\multicolumn{3}{c}{1010}&\multicolumn{3}{c}{  -5.65}&\multicolumn{3}{c}{1640} \\

&&&&&\multicolumn{2}{c}{48.4}&\multicolumn{2}{c}{45.6}&\multicolumn{2}{c}{32.1}&\multicolumn{2}{c}{48.3}&\multicolumn{2}{c}{45.5}&\multicolumn{2}{c}{33.0}&\multicolumn{2}{c}{48.4}&\multicolumn{2}{c}{45.5}&\multicolumn{2}{c}{32.7}&\multicolumn{2}{c}{48.4}&\multicolumn{2}{c}{45.4}&\multicolumn{2}{c}{33.6}&\multicolumn{2}{c}{48.4}&\multicolumn{2}{c}{45.7}&\multicolumn{2}{c}{...} \\

 OB\#21&   26.3&2.99&26.3& B0I&\multicolumn{3}{c}{  -6.08}&\multicolumn{3}{c}{1400}&\multicolumn{3}{c}{  -6.64}&\multicolumn{3}{c}{1130}&\multicolumn{3}{c}{  -6.40}&\multicolumn{3}{c}{1240}&\multicolumn{3}{c}{  -7.12}&\multicolumn{3}{c}{ 940}&\multicolumn{3}{c}{  -5.84}&\multicolumn{3}{c}{1530} \\

&&&&&\multicolumn{2}{c}{47.8}&\multicolumn{2}{c}{45.2}&\multicolumn{2}{c}{...}&\multicolumn{2}{c}{47.5}&\multicolumn{2}{c}{44.8}&\multicolumn{2}{c}{31.8}&\multicolumn{2}{c}{47.6}&\multicolumn{2}{c}{45.0}&\multicolumn{2}{c}{31.5}&\multicolumn{2}{c}{47.6}&\multicolumn{2}{c}{44.6}&\multicolumn{2}{c}{32.2}&\multicolumn{2}{c}{47.8}&\multicolumn{2}{c}{45.3}&\multicolumn{2}{c}{...} \\

OB\#22&   25.0&2.95&26.9&B0.5I&\multicolumn{3}{c}{  -6.11}&\multicolumn{3}{c}{1200}&\multicolumn{3}{c}{  -6.67}&\multicolumn{3}{c}{ 970}&\multicolumn{3}{c}{  -6.43}&\multicolumn{3}{c}{1060}&\multicolumn{3}{c}{  -7.15}&\multicolumn{3}{c}{ 810}&\multicolumn{3}{c}{  -5.87}&\multicolumn{3}{c}{1310} \\

&&&&&\multicolumn{2}{c}{47.6}&\multicolumn{2}{c}{44.9}&\multicolumn{2}{c}{...}&\multicolumn{2}{c}{47.1}&\multicolumn{2}{c}{44.2}&\multicolumn{2}{c}{31.5}&\multicolumn{2}{c}{47.4}&\multicolumn{2}{c}{44.6}&\multicolumn{2}{c}{...}&\multicolumn{2}{c}{47.1}&\multicolumn{2}{c}{44.2}&\multicolumn{2}{c}{...}&\multicolumn{2}{c}{47.7}&\multicolumn{2}{c}{45.1}&\multicolumn{2}{c}{...} \\ 
\hline\\
\noalign{
Notes: Mass loss rates $\dot{M}$ and terminal velocities $v_{\infty}$ are in units of
M$_{\odot}$\,yr$^{-1}$ and km\,s$^{-1}$ respectively. $Q_0, Q_1$ and $Q_2$ are on
a log scale and represent the photon luminosity (s$^{-1}$) in the
H\one, He\one\ and He\two\ ionizing continua.}
\end{tabular}
\end{sidewaystable*}
\subsection{O star model atmospheres}\label{Ogrid}

The current status of modelling O stars is reviewed by Kudritzki \&
Puls (2000).  To calculate realistic ionizing fluxes, the effects of
line blocking and blanketing have to be included in any expanding
non-LTE model atmosphere code. This makes the calculation complex and
computer-intensive because in an expanding atmosphere, the frequency
range that can be blocked by a single spectral line is increased, and
thus thousands of spectral lines from different ions overlap. Schaerer
\& de Koter (1997) used a Monte-Carlo approach (Schmutz 1991) which
has the advantage that the influence of millions of lines on the
emergent spectrum can be calculated, but the disadvantage that line
broadening is not included, and thus the line blanketing in the low velocity
part of the wind is underestimated.
The WM-basic code of Pauldrach et al. (2001) overcomes
these drawbacks by using an opacity sampling technique and a
consistent treatment of line blocking and blanketing to produce
synthetic high resolution spectra.  We have therefore used the
WM-basic code of Pauldrach et al. (2001) because of its fuller
treatment of line blanketing. We discuss the differences in the
predicted ionizing fluxes of the CoStar and WM-basic codes in 
Section~\ref{Ocomp}.

\begin{figure}
\psfig{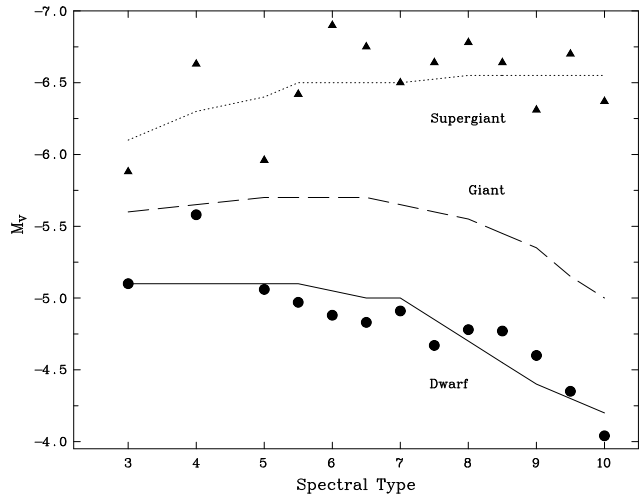}
\caption{The adopted absolute magnitude $M_V$ -- spectral type calibrations 
for dwarf (solid), giant (dashed) and supergiant (dotted) O stars.
The symbols indicate the mean values for each spectral type for dwarfs
and supergiants. The typical error on these values is $\pm 0.5$ mag.}
\label{Mv}
\end{figure}

In Tables 1 and 2, we present the model grid we have computed for O
and early B dwarfs and supergiants.  We defined the fundamental
parameters (effective temperature $T_{\rm {eff}}$; gravity $\log g$;
and photospheric radius $R_*$) of the stellar atmospheres using the
following approach. A revised absolute visual magnitude, $M_{\rm V}$--
spectral type scale for O dwarfs and supergiants was determined for
Galactic and LMC stars, following the approach and references cited by
Crowther \& Dessart (1998) in their study of early O stars.  A fit was
made to individual absolute magnitudes as illustrated in
Fig.~\ref{Mv}.

Next, we adopted the revised effective temperature vs. spectral type
calibration of Crowther (1998) for supergiants and dwarfs, which was
updated relative to Vacca, Garmany \& Shull (1996) to take account of
more recent hydrostatic results which neglect stellar winds and line
blanketing. There is increasing evidence that the temperature scale
for O dwarfs and especially supergiants based on hydrostatic models
is too high (Martins, Schaerer \& Hillier 2002; Crowther et
al. 2002b). Fortunately, the effect for O dwarfs is relatively small
$\sim5$\,kK, but future modifications will certainly be necessary.

We defined an effective temperature grid to span the domain of O and
early B stars from 25\,000--51\,000\,K, spaced at typically 0.03 dex
intervals to give 11 models covering the spectral classes from O3 to
B1.5. Using the $T_{\rm eff}$-bolometric correction calibration of
Vacca et al.  (1996) we determined bolometric magnitudes and
photospheric radii for each model. Since a discrepancy remains between
masses derived from spectroscopic and evolutionary models (Herrero et
al. 1992), we adopted (logarithmic) mean surface gravities from Vacca
et al. in all cases.

The appropriate terminal velocity $v_\infty$ for each model at solar
metallicity was determined using the $v_\infty$--spectral type
calibrations of Prinja, Barlow \& Howarth (1990) and Lamers, Snow \&
Lindholm (1995). The mass loss rate $\dot M$ was derived from the wind
momentum-luminosity relationship given by Kudritzki \& Puls (2000)
\[ \log \left[ \dot M v_\infty \left(R_*/R_\odot\right)^{0.5}\right]
= \log D_0 + x \log \frac{L}{L_\odot} \]
where $\log D_0$ and $x= 20.69, 1.51$ (O\,I); 19.87, 1.57 (O\,V); and
21.24, 1.34 (early B\,I). Since O star winds are driven through the
absorption of photospheric photons by extreme ultraviolet (EUV) metal
lines, mass loss rates and terminal velocities must depend on
metallicity.  Radiatively-driven wind theory (e.g. Kudritzki, Pauldrach
\& Puls 1987)
predicts that mass loss should scale with metallicity with an exponent
of 0.5--1.0. Recently Vink, de Koter \& Lamers (2001) predict an
exponent of 0.7 using Monte Carlo modelling. We have chosen to follow
the prescription of Leitherer, Robert \& Drissen (1992) in scaling
both $\dot M$ and $v_\infty$ with metallicity by using power law
exponents of 0.8 and 0.13 respectively i.e.
\[ \dot M_Z = Z^{0.8}\, \dot M_\odot;\ \ \ \ \
v_{\infty,Z} = Z^{0.13}\,v_{\infty,\odot}\]
The scaled values for $\dot M$ and $v_\infty$ are given for each
metallicity in Tables 1 and 2. 
\begin{sidewaystable*}
\label{tab_WN}
\begin{tabular}{@{}*{4}{c}*{6}{c}*{6}{c}*{6}{c}*{6}{c}*{6}{c}}
\noalign{{\bf Table 3:} WN Model Atmosphere Grid}\\
\hline
&&&&
\multicolumn{6}{c}{ \bf{Z$_{\odot}$}} &
\multicolumn{6}{c}{\bf{0.2\,Z$_{\odot}$}} &
\multicolumn{6}{c}{\bf{0.4\,Z$_{\odot}$}} &
\multicolumn{6}{c}{\bf{0.05\,Z$_{\odot}$}} &
\multicolumn{6}{c}{\bf{2.0\,Z$_{\odot}$}} \\ 
Model & $T_{\star}$ & $R_{\star}$ & $Y$ & \multicolumn{3}{c}{log
$\dot{M}$}&\multicolumn{3}{c}{$v_{\infty}$}&\multicolumn{3}{c}{log
$\dot{M}$}&\multicolumn{3}{c}{$v_{\infty}$}&\multicolumn{3}{c}{log
$\dot{M}$}&\multicolumn{3}{c}{$v_{\infty}$}&\multicolumn{3}{c}{log
$\dot{M}$}&\multicolumn{3}{c}{$v_{\infty}$}&\multicolumn{3}{c}{log
$\dot{M}$}&\multicolumn{3}{c}{$v_{\infty}$}\\
Ref. &(kK) & (R$_{\odot}$)&&
\multicolumn{2}{c}{ Q$_{0}$}&\multicolumn{2}{c}{
  Q$_{1}$}&\multicolumn{2}{c}{ Q$_{2}$}&\multicolumn{2}{c}{
  Q$_{0}$}&\multicolumn{2}{c}{ Q$_{1}$}&\multicolumn{2}{c}{
  Q$_{2}$}&\multicolumn{2}{c}{ Q$_{0}$}&\multicolumn{2}{c}{
  Q$_{1}$}&\multicolumn{2}{c}{ Q$_{2}$}&\multicolumn{2}{c}{
  Q$_{0}$}&\multicolumn{2}{c}{ Q$_{1}$}&\multicolumn{2}{c}{
  Q$_{2}$}&\multicolumn{2}{c}{ Q$_{0}$}&\multicolumn{2}{c}{
  Q$_{1}$}&\multicolumn{2}{c}{ Q$_{2}$}\\
\hline
WN\#1  &   30.0&20.3&0.76&\multicolumn{3}{c}{  -4.93}&\multicolumn{3}{c}{1100}&\multicolumn{3}{c}{  -5.49}&\multicolumn{3}{c}{ 890}&\multicolumn{3}{c}{  -5.25}&\multicolumn{3}{c}{ 970}&\multicolumn{3}{c}{  -5.97}&\multicolumn{3}{c}{ 740}&\multicolumn{3}{c}{  -4.69}&\multicolumn{3}{c}{1200} \\
&&&&\multicolumn{2}{c}{48.8}&\multicolumn{2}{c}{45.0}&\multicolumn{2}{c}{29.5}&\multicolumn{2}{c}{48.8}&\multicolumn{2}{c}{46.8}&\multicolumn{2}{c}{34.7}&\multicolumn{2}{c}{48.8}&\multicolumn{2}{c}{46.6}&\multicolumn{2}{c}{34.4}&\multicolumn{2}{c}{48.8}&\multicolumn{2}{c}{46.8}&\multicolumn{2}{c}{35.3}&\multicolumn{2}{c}{48.7}&\multicolumn{2}{c}{44.3}&\multicolumn{2}{c}{28.0} \\
WN\#2  &   32.5&17.2&0.78&\multicolumn{3}{c}{  -4.91}&\multicolumn{3}{c}{1170}&\multicolumn{3}{c}{  -5.47}&\multicolumn{3}{c}{ 940}&\multicolumn{3}{c}{  -5.23}&\multicolumn{3}{c}{1030}&\multicolumn{3}{c}{  -5.95}&\multicolumn{3}{c}{ 790}&\multicolumn{3}{c}{  -4.67}&\multicolumn{3}{c}{1280} \\
&&&&\multicolumn{2}{c}{49.0}&\multicolumn{2}{c}{47.4}&\multicolumn{2}{c}{34.3}&\multicolumn{2}{c}{49.0}&\multicolumn{2}{c}{47.7}&\multicolumn{2}{c}{37.0}&\multicolumn{2}{c}{49.0}&\multicolumn{2}{c}{47.7}&\multicolumn{2}{c}{36.8}&\multicolumn{2}{c}{49.0}&\multicolumn{2}{c}{47.7}&\multicolumn{2}{c}{36.5}&\multicolumn{2}{c}{48.8}&\multicolumn{2}{c}{44.3}&\multicolumn{2}{c}{28.3} \\
WN\#3  &   35.0&14.9&0.80&\multicolumn{3}{c}{  -4.89}&\multicolumn{3}{c}{1240}&\multicolumn{3}{c}{  -5.45}&\multicolumn{3}{c}{1000}&\multicolumn{3}{c}{  -5.21}&\multicolumn{3}{c}{1100}&\multicolumn{3}{c}{  -5.93}&\multicolumn{3}{c}{ 840}&\multicolumn{3}{c}{  -4.65}&\multicolumn{3}{c}{1350} \\
&&&&\multicolumn{2}{c}{49.1}&\multicolumn{2}{c}{47.8}&\multicolumn{2}{c}{37.1}&\multicolumn{2}{c}{49.1}&\multicolumn{2}{c}{48.1}&\multicolumn{2}{c}{37.3}&\multicolumn{2}{c}{49.1}&\multicolumn{2}{c}{48.1}&\multicolumn{2}{c}{37.6}&\multicolumn{2}{c}{49.1}&\multicolumn{2}{c}{48.2}&\multicolumn{2}{c}{36.9}&\multicolumn{2}{c}{49.0}&\multicolumn{2}{c}{47.1}&\multicolumn{2}{c}{34.5} \\
WN\#4  &   40.0&11.4&0.85&\multicolumn{3}{c}{  -4.83}&\multicolumn{3}{c}{1370}&\multicolumn{3}{c}{  -5.39}&\multicolumn{3}{c}{1110}&\multicolumn{3}{c}{  -5.15}&\multicolumn{3}{c}{1210}&\multicolumn{3}{c}{  -5.87}&\multicolumn{3}{c}{ 920}&\multicolumn{3}{c}{  -4.59}&\multicolumn{3}{c}{1490} \\
&&&&\multicolumn{2}{c}{49.2}&\multicolumn{2}{c}{48.1}&\multicolumn{2}{c}{38.1}&\multicolumn{2}{c}{49.2}&\multicolumn{2}{c}{48.4}&\multicolumn{2}{c}{37.7}&\multicolumn{2}{c}{49.2}&\multicolumn{2}{c}{48.3}&\multicolumn{2}{c}{37.7}&\multicolumn{2}{c}{49.2}&\multicolumn{2}{c}{48.5}&\multicolumn{2}{c}{37.7}&\multicolumn{2}{c}{49.2}&\multicolumn{2}{c}{47.7}&\multicolumn{2}{c}{34.3} \\
WN\#5  &   45.0& 9.0&0.91&\multicolumn{3}{c}{  -4.76}&\multicolumn{3}{c}{1520}&\multicolumn{3}{c}{  -5.32}&\multicolumn{3}{c}{1230}&\multicolumn{3}{c}{  -5.08}&\multicolumn{3}{c}{1340}&\multicolumn{3}{c}{  -5.80}&\multicolumn{3}{c}{1020}&\multicolumn{3}{c}{  -4.52}&\multicolumn{3}{c}{1660} \\
&&&&\multicolumn{2}{c}{49.3}&\multicolumn{2}{c}{48.4}&\multicolumn{2}{c}{38.1}&\multicolumn{2}{c}{49.3}&\multicolumn{2}{c}{48.6}&\multicolumn{2}{c}{38.2}&\multicolumn{2}{c}{49.3}&\multicolumn{2}{c}{48.6}&\multicolumn{2}{c}{38.1}&\multicolumn{2}{c}{49.3}&\multicolumn{2}{c}{48.7}&\multicolumn{2}{c}{36.9}&\multicolumn{2}{c}{49.2}&\multicolumn{2}{c}{47.9}&\multicolumn{2}{c}{35.2} \\
WN\#6  &   50.0& 7.3&0.94&\multicolumn{3}{c}{  -4.73}&\multicolumn{3}{c}{1670}&\multicolumn{3}{c}{  -5.29}&\multicolumn{3}{c}{1350}&\multicolumn{3}{c}{  -5.05}&\multicolumn{3}{c}{1480}&\multicolumn{3}{c}{  -5.77}&\multicolumn{3}{c}{1130}&\multicolumn{3}{c}{  -4.49}&\multicolumn{3}{c}{1820} \\
&&&&\multicolumn{2}{c}{49.3}&\multicolumn{2}{c}{48.6}&\multicolumn{2}{c}{38.2}&\multicolumn{2}{c}{49.3}&\multicolumn{2}{c}{48.8}&\multicolumn{2}{c}{38.6}&\multicolumn{2}{c}{49.3}&\multicolumn{2}{c}{48.8}&\multicolumn{2}{c}{38.6}&\multicolumn{2}{c}{49.3}&\multicolumn{2}{c}{48.8}&\multicolumn{2}{c}{38.5}&\multicolumn{2}{c}{49.3}&\multicolumn{2}{c}{48.3}&\multicolumn{2}{c}{38.5} \\
WN\#7  &   60.0& 5.1&0.98&\multicolumn{3}{c}{  -4.69}&\multicolumn{3}{c}{1960}&\multicolumn{3}{c}{  -5.25}&\multicolumn{3}{c}{1580}&\multicolumn{3}{c}{  -5.01}&\multicolumn{3}{c}{1730}&\multicolumn{3}{c}{  -5.73}&\multicolumn{3}{c}{1320}&\multicolumn{3}{c}{  -4.45}&\multicolumn{3}{c}{2140} \\
&&&&\multicolumn{2}{c}{49.4}&\multicolumn{2}{c}{48.9}&\multicolumn{2}{c}{38.4}&\multicolumn{2}{c}{49.4}&\multicolumn{2}{c}{49.0}&\multicolumn{2}{c}{39.1}&\multicolumn{2}{c}{49.4}&\multicolumn{2}{c}{49.0}&\multicolumn{2}{c}{39.2}&\multicolumn{2}{c}{49.4}&\multicolumn{2}{c}{49.0}&\multicolumn{2}{c}{38.9}&\multicolumn{2}{c}{49.3}&\multicolumn{2}{c}{48.6}&\multicolumn{2}{c}{38.6} \\
WN\#8  &   70.0& 3.7&0.98&\multicolumn{3}{c}{  -4.69}&\multicolumn{3}{c}{2270}&\multicolumn{3}{c}{  -5.25}&\multicolumn{3}{c}{1840}&\multicolumn{3}{c}{  -5.01}&\multicolumn{3}{c}{2010}&\multicolumn{3}{c}{  -5.73}&\multicolumn{3}{c}{1530}&\multicolumn{3}{c}{  -4.45}&\multicolumn{3}{c}{2480} \\
&&&&\multicolumn{2}{c}{49.4}&\multicolumn{2}{c}{49.0}&\multicolumn{2}{c}{39.6}&\multicolumn{2}{c}{49.4}&\multicolumn{2}{c}{49.1}&\multicolumn{2}{c}{39.8}&\multicolumn{2}{c}{49.4}&\multicolumn{2}{c}{49.1}&\multicolumn{2}{c}{39.8}&\multicolumn{2}{c}{49.3}&\multicolumn{2}{c}{49.0}&\multicolumn{2}{c}{39.8}&\multicolumn{2}{c}{49.4}&\multicolumn{2}{c}{48.8}&\multicolumn{2}{c}{38.6} \\
WN\#9  &   80.0& 2.8&0.98&\multicolumn{3}{c}{  -4.69}&\multicolumn{3}{c}{2550}&\multicolumn{3}{c}{  -5.25}&\multicolumn{3}{c}{2060}&\multicolumn{3}{c}{  -5.01}&\multicolumn{3}{c}{2260}&\multicolumn{3}{c}{  -5.73}&\multicolumn{3}{c}{1720}&\multicolumn{3}{c}{  -4.45}&\multicolumn{3}{c}{2790} \\
&&&&\multicolumn{2}{c}{49.4}&\multicolumn{2}{c}{49.1}&\multicolumn{2}{c}{40.1}&\multicolumn{2}{c}{49.3}&\multicolumn{2}{c}{49.1}&\multicolumn{2}{c}{40.4}&\multicolumn{2}{c}{49.4}&\multicolumn{2}{c}{49.1}&\multicolumn{2}{c}{40.3}&\multicolumn{2}{c}{49.3}&\multicolumn{2}{c}{49.1}&\multicolumn{2}{c}{47.5}&\multicolumn{2}{c}{49.4}&\multicolumn{2}{c}{48.9}&\multicolumn{2}{c}{38.9} \\
WN\#10  &   90.0& 2.3&0.98&\multicolumn{3}{c}{  -4.69}&\multicolumn{3}{c}{2850}&\multicolumn{3}{c}{  -5.25}&\multicolumn{3}{c}{2310}&\multicolumn{3}{c}{  -5.01}&\multicolumn{3}{c}{2520}&\multicolumn{3}{c}{  -5.73}&\multicolumn{3}{c}{1930}&\multicolumn{3}{c}{  -4.45}&\multicolumn{3}{c}{3110} \\
&&&&\multicolumn{2}{c}{49.3}&\multicolumn{2}{c}{49.1}&\multicolumn{2}{c}{40.6}&\multicolumn{2}{c}{49.3}&\multicolumn{2}{c}{49.1}&\multicolumn{2}{c}{47.9}&\multicolumn{2}{c}{49.3}&\multicolumn{2}{c}{49.1}&\multicolumn{2}{c}{40.4}&\multicolumn{2}{c}{49.3}&\multicolumn{2}{c}{49.1}&\multicolumn{2}{c}{48.0}&\multicolumn{2}{c}{49.3}&\multicolumn{2}{c}{49.0}&\multicolumn{2}{c}{39.5} \\
WN\#11  &  100.0& 1.8&0.98&\multicolumn{3}{c}{  -4.69}&\multicolumn{3}{c}{3120}&\multicolumn{3}{c}{  -5.25}&\multicolumn{3}{c}{2530}&\multicolumn{3}{c}{  -5.01}&\multicolumn{3}{c}{2760}&\multicolumn{3}{c}{  -5.73}&\multicolumn{3}{c}{2110}&\multicolumn{3}{c}{  -4.45}&\multicolumn{3}{c}{3410} \\
&&&&\multicolumn{2}{c}{49.3}&\multicolumn{2}{c}{49.1}&\multicolumn{2}{c}{41.2}&\multicolumn{2}{c}{49.3}&\multicolumn{2}{c}{49.1}&\multicolumn{2}{c}{48.3}&\multicolumn{2}{c}{49.3}&\multicolumn{2}{c}{49.1}&\multicolumn{2}{c}{48.1}&\multicolumn{2}{c}{49.3}&\multicolumn{2}{c}{49.1}&\multicolumn{2}{c}{48.3}&\multicolumn{2}{c}{49.3}&\multicolumn{2}{c}{49.0}&\multicolumn{2}{c}{40.4} \\ 
WN\#12  &  120.0& 1.3&0.98&\multicolumn{3}{c}{  -4.69}&\multicolumn{3}{c}{3700}&\multicolumn{3}{c}{  -5.25}&\multicolumn{3}{c}{3000}&\multicolumn{3}{c}{  -5.01}&\multicolumn{3}{c}{3300}&\multicolumn{3}{c}{  -5.73}&\multicolumn{3}{c}{2500}&\multicolumn{3}{c}{  -4.45}&\multicolumn{3}{c}{4050} \\
&&&&\multicolumn{2}{c}{49.3}&\multicolumn{2}{c}{49.1}&\multicolumn{2}{c}{49.2}&\multicolumn{2}{c}{49.2}&\multicolumn{2}{c}{49.1}&\multicolumn{2}{c}{48.5}&\multicolumn{2}{c}{49.3}&\multicolumn{2}{c}{49.1}&\multicolumn{2}{c}{48.5}&\multicolumn{2}{c}{49.2}&\multicolumn{2}{c}{49.1}&\multicolumn{2}{c}{48.6}&\multicolumn{2}{c}{49.3}&\multicolumn{2}{c}{49.1}&\multicolumn{2}{c}{41.1} \\      
\hline\\
\noalign{
Notes: Mass loss rates $\dot{M}$ and terminal velocities $v_{\infty}$ are in units of
M$_{\odot}$\,yr$^{-1}$ and km\,s$^{-1}$ respectively. $Q_0, Q_1$ and $Q_2$ are on
a log scale and represent the photon luminosity (s$^{-1}$) in the
H\one, He\one\ and He\two\ ionizing continua.}
\end{tabular}
\end{sidewaystable*}
\begin{sidewaystable*}
\label{tab_WC}
\begin{tabular}{@{}*{4}{c}*{6}{c}*{6}{c}*{6}{c}*{6}{c}*{6}{c}}
\noalign{{\bf Table 4:} WC Model Atmosphere Grid}\\
\hline
&&&&
\multicolumn{6}{c}{ \bf{Z$_{\odot}$}} &
\multicolumn{6}{c}{\bf{0.2\,Z$_{\odot}$}} &
\multicolumn{6}{c}{\bf{0.4\,Z$_{\odot}$}} &
\multicolumn{6}{c}{\bf{0.05\,Z$_{\odot}$}} &
\multicolumn{6}{c}{\bf{2.0\,Z$_{\odot}$}} \\ 
Model & $T_{\star}$ & $R_{\star}$ & $Y$ & \multicolumn{3}{c}{log
$\dot{M}$}&\multicolumn{3}{c}{$v_{\infty}$}&\multicolumn{3}{c}{log
$\dot{M}$}&\multicolumn{3}{c}{$v_{\infty}$}&\multicolumn{3}{c}{log
$\dot{M}$}&\multicolumn{3}{c}{$v_{\infty}$}&\multicolumn{3}{c}{log
$\dot{M}$}&\multicolumn{3}{c}{$v_{\infty}$}&\multicolumn{3}{c}{log
$\dot{M}$}&\multicolumn{3}{c}{$v_{\infty}$}\\
Ref. &(kK) & (R$_{\odot}$)&&
\multicolumn{2}{c}{ Q$_{0}$}&\multicolumn{2}{c}{
  Q$_{1}$}&\multicolumn{2}{c}{ Q$_{2}$}&\multicolumn{2}{c}{
  Q$_{0}$}&\multicolumn{2}{c}{ Q$_{1}$}&\multicolumn{2}{c}{
  Q$_{2}$}&\multicolumn{2}{c}{ Q$_{0}$}&\multicolumn{2}{c}{
  Q$_{1}$}&\multicolumn{2}{c}{ Q$_{2}$}&\multicolumn{2}{c}{
  Q$_{0}$}&\multicolumn{2}{c}{ Q$_{1}$}&\multicolumn{2}{c}{
  Q$_{2}$}&\multicolumn{2}{c}{ Q$_{0}$}&\multicolumn{2}{c}{
  Q$_{1}$}&\multicolumn{2}{c}{ Q$_{2}$}\\
\hline
WC\#1  &   40.0& 9.3&0.56&\multicolumn{3}{c}{  -4.72}&\multicolumn{3}{c}{1250}&
\multicolumn{3}{c}{  -5.28}&\multicolumn{3}{c}{1010}&\multicolumn{3}{c}{  -5.04}&
\multicolumn{3}{c}{1100}&\multicolumn{3}{c}{  -5.76}&\multicolumn{3}{c}{ 840}&
\multicolumn{3}{c}{  -4.48}&\multicolumn{3}{c}{1360}\\
&&&&\multicolumn{2}{c}{48.9}&\multicolumn{2}{c}{41.8}&\multicolumn{2}{c}{26.1}&
\multicolumn{2}{c}{49.0}&\multicolumn{2}{c}{47.3}&\multicolumn{2}{c}{33.0}&
\multicolumn{2}{c}{49.0}&\multicolumn{2}{c}{47.6}&\multicolumn{2}{c}{34.0}&
\multicolumn{2}{c}{48.9}&\multicolumn{2}{c}{46.9}&\multicolumn{2}{c}{31.4}&
\multicolumn{2}{c}{48.7}&\multicolumn{2}{c}{42.1}&\multicolumn{2}{c}{25.9}\\
WC\#2  &   43.0& 8.0&0.56&\multicolumn{3}{c}{  -4.72}&\multicolumn{3}{c}{1380}&
\multicolumn{3}{c}{  -5.28}&\multicolumn{3}{c}{1110}&\multicolumn{3}{c}{  -5.04}&
\multicolumn{3}{c}{1220}&\multicolumn{3}{c}{  -5.76}&\multicolumn{3}{c}{ 930}&
\multicolumn{3}{c}{  -4.48}&\multicolumn{3}{c}{1510} \\
&&&&\multicolumn{2}{c}{49.0}&\multicolumn{2}{c}{42.1}&\multicolumn{2}{c}{27.5}&
\multicolumn{2}{c}{49.1}&\multicolumn{2}{c}{47.7}&\multicolumn{2}{c}{34.8}&
\multicolumn{2}{c}{49.1}&\multicolumn{2}{c}{47.9}&\multicolumn{2}{c}{35.4}&
\multicolumn{2}{c}{49.0}&\multicolumn{2}{c}{47.5}&\multicolumn{2}{c}{33.4}&
\multicolumn{2}{c}{48.8}&\multicolumn{2}{c}{39.6}&\multicolumn{2}{c}{23.8} \\
WC\#3  &   46.0& 7.0&0.56&\multicolumn{3}{c}{  -4.72}&\multicolumn{3}{c}{1510}&
\multicolumn{3}{c}{  -5.28}&\multicolumn{3}{c}{1220}&\multicolumn{3}{c}{  -5.04}&
\multicolumn{3}{c}{1340}&\multicolumn{3}{c}{  -5.76}&\multicolumn{3}{c}{1020}&
\multicolumn{3}{c}{  -4.48}&\multicolumn{3}{c}{1650} \\
&&&&\multicolumn{2}{c}{49.0}&\multicolumn{2}{c}{46.7}&\multicolumn{2}{c}{31.3}&
\multicolumn{2}{c}{49.1}&\multicolumn{2}{c}{48.0}&\multicolumn{2}{c}{36.1}&
\multicolumn{2}{c}{49.1}&\multicolumn{2}{c}{47.9}&\multicolumn{2}{c}{34.8}&
\multicolumn{2}{c}{49.1}&\multicolumn{2}{c}{48.1}&\multicolumn{2}{c}{36.1}&
\multicolumn{2}{c}{48.9}&\multicolumn{2}{c}{42.1}&\multicolumn{2}{c}{26.6} \\
WC\#4  &   50.0& 5.9&0.56&\multicolumn{3}{c}{  -4.72}&\multicolumn{3}{c}{1650}&
\multicolumn{3}{c}{  -5.28}&\multicolumn{3}{c}{1330}&\multicolumn{3}{c}{  -5.04}&
\multicolumn{3}{c}{1460}&\multicolumn{3}{c}{  -5.76}&\multicolumn{3}{c}{1110}&
\multicolumn{3}{c}{  -4.48}&\multicolumn{3}{c}{1800} \\
&&&&\multicolumn{2}{c}{49.1}&\multicolumn{2}{c}{47.5}&\multicolumn{2}{c}{33.1}&
\multicolumn{2}{c}{49.1}&\multicolumn{2}{c}{48.3}&\multicolumn{2}{c}{37.0}&
\multicolumn{2}{c}{49.1}&\multicolumn{2}{c}{48.1}&\multicolumn{2}{c}{36.6}&
\multicolumn{2}{c}{49.1}&\multicolumn{2}{c}{48.4}&\multicolumn{2}{c}{35.3}&
\multicolumn{2}{c}{49.0}&\multicolumn{2}{c}{42.2}&\multicolumn{2}{c}{27.3} \\
WC\#5  &   55.0& 4.9&0.56&\multicolumn{3}{c}{  -4.72}&\multicolumn{3}{c}{1870}&
\multicolumn{3}{c}{  -5.28}&\multicolumn{3}{c}{1510}&\multicolumn{3}{c}{  -5.04}&
\multicolumn{3}{c}{1660}&\multicolumn{3}{c}{  -5.76}&\multicolumn{3}{c}{1260}&
\multicolumn{3}{c}{  -4.48}&\multicolumn{3}{c}{2040} \\
&&&&\multicolumn{2}{c}{49.1}&\multicolumn{2}{c}{47.9}&\multicolumn{2}{c}{31.3}&
\multicolumn{2}{c}{49.2}&\multicolumn{2}{c}{48.5}&\multicolumn{2}{c}{37.3}&
\multicolumn{2}{c}{49.2}&\multicolumn{2}{c}{48.4}&\multicolumn{2}{c}{37.4}&
\multicolumn{2}{c}{49.2}&\multicolumn{2}{c}{48.6}&\multicolumn{2}{c}{40.3}&
\multicolumn{2}{c}{49.0}&\multicolumn{2}{c}{42.4}&\multicolumn{2}{c}{28.5} \\
WC\#6  &   60.0& 4.1&0.56&\multicolumn{3}{c}{  -4.72}&\multicolumn{3}{c}{2100}&
\multicolumn{3}{c}{  -5.28}&\multicolumn{3}{c}{1700}&\multicolumn{3}{c}{  -5.04}&
\multicolumn{3}{c}{1860}&\multicolumn{3}{c}{  -5.76}&\multicolumn{3}{c}{1420}&
\multicolumn{3}{c}{  -4.48}&\multicolumn{3}{c}{2290} \\
&&&&\multicolumn{2}{c}{49.1}&\multicolumn{2}{c}{48.1}&\multicolumn{2}{c}{36.7}&
\multicolumn{2}{c}{49.2}&\multicolumn{2}{c}{48.7}&\multicolumn{2}{c}{37.6}&
\multicolumn{2}{c}{49.2}&\multicolumn{2}{c}{48.6}&\multicolumn{2}{c}{37.5}&
\multicolumn{2}{c}{49.2}&\multicolumn{2}{c}{48.7}&\multicolumn{2}{c}{39.9}&
\multicolumn{2}{c}{49.0}&\multicolumn{2}{c}{47.9}&\multicolumn{2}{c}{34.8} \\
WC\#7  &   70.0& 3.0&0.56&\multicolumn{3}{c}{  -4.72}&\multicolumn{3}{c}{2450}&
\multicolumn{3}{c}{  -5.28}&\multicolumn{3}{c}{1980}&\multicolumn{3}{c}{  -5.04}&
\multicolumn{3}{c}{2170}&\multicolumn{3}{c}{  -5.76}&\multicolumn{3}{c}{1650}&
\multicolumn{3}{c}{  -4.48}&\multicolumn{3}{c}{2680} \\
&&&&\multicolumn{2}{c}{49.1}&\multicolumn{2}{c}{48.5}&\multicolumn{2}{c}{37.9}&
\multicolumn{2}{c}{49.2}&\multicolumn{2}{c}{48.8}&\multicolumn{2}{c}{38.3}&
\multicolumn{2}{c}{49.2}&\multicolumn{2}{c}{48.7}&\multicolumn{2}{c}{37.9}&
\multicolumn{2}{c}{49.2}&\multicolumn{2}{c}{48.8}&\multicolumn{2}{c}{40.3}&
\multicolumn{2}{c}{49.1}&\multicolumn{2}{c}{47.9}&\multicolumn{2}{c}{32.9} \\
WC\#8  &   80.0& 2.3&0.56&\multicolumn{3}{c}{  -4.72}&\multicolumn{3}{c}{2800}&
\multicolumn{3}{c}{  -5.28}&\multicolumn{3}{c}{2270}&\multicolumn{3}{c}{  -5.04}&
\multicolumn{3}{c}{2480}&\multicolumn{3}{c}{  -5.76}&\multicolumn{3}{c}{1890}&
\multicolumn{3}{c}{  -4.48}&\multicolumn{3}{c}{3060} \\
&&&&\multicolumn{2}{c}{49.2}&\multicolumn{2}{c}{48.6}&\multicolumn{2}{c}{38.2}&
\multicolumn{2}{c}{49.2}&\multicolumn{2}{c}{48.9}&\multicolumn{2}{c}{39.9}&
\multicolumn{2}{c}{49.2}&\multicolumn{2}{c}{48.8}&\multicolumn{2}{c}{38.6}&
\multicolumn{2}{c}{49.2}&\multicolumn{2}{c}{48.9}&\multicolumn{2}{c}{40.3}&
\multicolumn{2}{c}{49.1}&\multicolumn{2}{c}{48.2}&\multicolumn{2}{c}{37.3} \\
WC\#9  &   90.0& 1.8&0.56&\multicolumn{3}{c}{  -4.72}&\multicolumn{3}{c}{3200}&
\multicolumn{3}{c}{  -5.28}&\multicolumn{3}{c}{2590}&\multicolumn{3}{c}{  -5.04}&
\multicolumn{3}{c}{2840}&\multicolumn{3}{c}{  -5.76}&\multicolumn{3}{c}{2160}&
\multicolumn{3}{c}{  -4.48}&\multicolumn{3}{c}{3500} \\
&&&&\multicolumn{2}{c}{49.2}&\multicolumn{2}{c}{48.7}&\multicolumn{2}{c}{38.4}&
\multicolumn{2}{c}{49.2}&\multicolumn{2}{c}{48.9}&\multicolumn{2}{c}{42.6}&
\multicolumn{2}{c}{49.2}&\multicolumn{2}{c}{48.9}&\multicolumn{2}{c}{39.3}&
\multicolumn{2}{c}{49.2}&\multicolumn{2}{c}{48.9}&\multicolumn{2}{c}{46.9}&
\multicolumn{2}{c}{49.1}&\multicolumn{2}{c}{48.4}&\multicolumn{2}{c}{38.3} \\
WC\#10  &  100.0& 1.5&0.56&\multicolumn{3}{c}{  -4.72}&\multicolumn{3}{c}{3600}&
\multicolumn{3}{c}{  -5.28}&\multicolumn{3}{c}{2920}&\multicolumn{3}{c}{  -5.04}&
\multicolumn{3}{c}{3190}&\multicolumn{3}{c}{  -5.76}&\multicolumn{3}{c}{2430}&
\multicolumn{3}{c}{  -4.48}&\multicolumn{3}{c}{3930} \\
&&&&\multicolumn{2}{c}{49.2}&\multicolumn{2}{c}{48.8}&\multicolumn{2}{c}{38.8}&
\multicolumn{2}{c}{49.2}&\multicolumn{2}{c}{48.9}&\multicolumn{2}{c}{40.7}&
\multicolumn{2}{c}{49.2}&\multicolumn{2}{c}{48.9}&\multicolumn{2}{c}{40.3}&
\multicolumn{2}{c}{49.1}&\multicolumn{2}{c}{48.9}&\multicolumn{2}{c}{47.7}&
\multicolumn{2}{c}{49.1}&\multicolumn{2}{c}{48.5}&\multicolumn{2}{c}{35.6} \\  
WC\#11  &  120.0& 1.0&0.56&\multicolumn{3}{c}{  -4.72}&\multicolumn{3}{c}{4400}&
\multicolumn{3}{c}{  -5.28}&\multicolumn{3}{c}{3570}&\multicolumn{3}{c}{  -5.04}&
\multicolumn{3}{c}{3900}&\multicolumn{3}{c}{  -5.76}&\multicolumn{3}{c}{3000}&
\multicolumn{3}{c}{  -4.48}&\multicolumn{3}{c}{4810} \\
&&&&\multicolumn{2}{c}{49.2}&\multicolumn{2}{c}{48.8}&\multicolumn{2}{c}{39.4}&
\multicolumn{2}{c}{49.1}&\multicolumn{2}{c}{48.9}&\multicolumn{2}{c}{48.1}&
\multicolumn{2}{c}{49.2}&\multicolumn{2}{c}{48.9}&\multicolumn{2}{c}{47.6}&
\multicolumn{2}{c}{49.1}&\multicolumn{2}{c}{48.9}&\multicolumn{2}{c}{48.2}&
\multicolumn{2}{c}{49.1}&\multicolumn{2}{c}{48.6}&\multicolumn{2}{c}{38.8} \\  
WC\#12  &  140.0& 0.8&0.56&\multicolumn{3}{c}{  -4.72}&\multicolumn{3}{c}{5200}&
\multicolumn{3}{c}{  -5.28}&\multicolumn{3}{c}{4220}&\multicolumn{3}{c}{  -5.04}&
\multicolumn{3}{c}{4620}&\multicolumn{3}{c}{  -5.76}&\multicolumn{3}{c}{3520}&
\multicolumn{3}{c}{  -4.48}&\multicolumn{3}{c}{5690} \\
&&&&\multicolumn{2}{c}{49.1}&\multicolumn{2}{c}{48.9}&\multicolumn{2}{c}{40.0}&
\multicolumn{2}{c}{49.1}&\multicolumn{2}{c}{48.9}&\multicolumn{2}{c}{48.8}&
\multicolumn{2}{c}{49.1}&\multicolumn{2}{c}{48.9}&\multicolumn{2}{c}{48.2}&
\multicolumn{2}{c}{49.0}&\multicolumn{2}{c}{48.9}&\multicolumn{2}{c}{48.4}&
\multicolumn{2}{c}{49.1}&\multicolumn{2}{c}{48.7}&\multicolumn{2}{c}{39.1} \\  
\hline\\
\noalign{
Notes: Mass loss rates $\dot{M}$ and terminal velocities $v_{\infty}$ are in units of
M$_{\odot}$\,yr$^{-1}$ and km\,s$^{-1}$ respectively. $Q_0, Q_1$ and $Q_2$ are on
a log scale and represent the photon luminosity (s$^{-1}$) in the
H\one, He\one\ and He\two\ ionizing continua.}
\end{tabular}
\end{sidewaystable*}
To compute the models, we used version 1.16 of the WM-basic code which
includes atomic models of all the important ions of 149 ionization
stages of 26 elements from H to Zn. The wind structure was solved for
consistently in the hydrodynamic part of WM-basic, corresponding to a
velocity law with exponent $\beta \sim 0.8$. Intervention was required
in order that the required mass loss rate and terminal velocity were
obtained, for the stellar input parameters, via ``force multipliers''
which describe the radiative line acceleration. We tuned the force
multipliers for each model by iteration to achieve the wind parameters
given in Tables 1 and 2 ($\dot M$ to within 5 per cent and $v_\infty$
to within $\pm 100$\, km\,s$^{-1}$).  We have not included the effect
of shocks in the stellar wind since shock characteristics are
ill-defined across the O spectral range. The presence of shocks will
increase the ionizing flux at EUV and X-ray wavelengths (see Pauldrach
et al. 2001 for details).  In Tables 1 and 2, we also provide the
photon luminosity in the H\one\ ($\log Q_0$), He\one\ ($\log Q_1$) and
He\two\ ($\log Q_2$) ionizing continua for each model.

\subsection{W-R star model atmospheres}\label{W-Rgrid}
The basic parameters of W-R stars have been substantially revised in
recent years, due to the availability of increasingly sophisticated
model atmospheres that are now able to reproduce their complex
emission line spectra (see e.g. reviews of Hillier (1999); Crowther
(1999)).

We have calculated a grid of expanding, non-LTE, line-blanketed WN and
WC model atmospheres with the {\sc cmfgen} code of Hillier \& Miller
(1998) using what we consider to be the most appropriate parameters
for these stars.  Line blanketing is accounted for in this code using
the concept of super-levels, with reasonably sophisticated model atoms
for the dominant ionization stages, namely H, He, C, N, O, Ne, Si, S,
Fe in WN stars and He, C, O, Si, Fe in WC stars. Subsequent to the
completion of this grid, computational improvements permitted many
other elements and ionization stages to be included in {\sc cmfgen},
but tests indicate that these tend to be second order effects.

In Tables 3 and 4, we present the parameters adopted in our WN and WC
grids, parameterized by the stellar temperature $T_*$, the stellar
radius $R_*$ (both defined at a Rosseland optical depth of $\sim 10$;
see Section~\ref{sb99} for a discussion of W-R temperature
definitions), and the helium mass fraction $Y$.  To define these
parameters, we used results from recent studies presenting
line-blanketed model atmosphere analyses of W-R stars.  For the WN
class, our main references were Herald, Hillier \& Schulte-Ladbeck
(2001) and Hillier \& Miller (1998), supplemented by Crowther \& Smith
(1997) for H/He ratios. For the WC class, we were guided by the
analyses of Hillier \& Miller (1999), Dessart et al. (2000) and
Crowther et al. (2002a). Nugis \& Lamers (2000) present a compilation
of the stellar parameters for 64 Galactic W-R stars and provide a good
discussion of the uncertainties.

We defined a $T_*$ range of 30\,000--120\,000\,K for the WNs and
assigned a decreasing hydrogen content from 24 per cent (representing
a cool late WN star) to 2 per cent (representing a hot WNE star with
$T_* > 60\,000$\,K), plus CNO equilibrium values in all cases. 
For the WC stars, we chose a $T_*$ range of
40\,000--140\,000\,K, and fixed C/He=0.2 and C/O=4 by number. We
assumed representative luminosities of $L=3 \times 10^5$\,L$_\odot$
and $2 \times 10^5$\,L$_\odot$ for the WN and WC models respectively.

The winds of W-R stars are believed to be inhomogeneous or clumped.
Evidence for this comes from observations of line profile variability
(e.g. Moffat et al. 1988) and the over-prediction of the strength of
the scattering wings in theoretical line profiles (Hillier 1991).  The
significance of clumping is that it leads to a downward revision in
W-R mass loss rates although the exact amount is unclear, since line
profile analyses give $\dot{M}/\sqrt f$ where $f$ is the volume
filling factor (e.g. Hillier \& Miller 1999). In an analysis of the
infrared and radio continuum fluxes for a large sample of W-R stars,
Nugis, Crowther \& Willis (1998) derived clumping-corrected mass loss
rates a factor of 3--5 lower than the smooth wind values. This study,
however, used a radio sample which is biased towards higher mass loss
rates, and thus the clumping factors should be viewed as upper limits
(C. Leitherer, priv. comm.).  Recent line profile analyses using {\sc
cmfgen} find good consistency using clumping factors of $f \sim 0.1$
(e.g. Dessart et al. 2000; Crowther et al. 2002a), which correspond to
mass loss rates that are a factor of $\sim 3$ times lower than
homogeneous models.

To derive representative mass loss rates at solar metallicity, we used
the mass loss--luminosity--chemical composition relationships given by
Nugis \& Lamers (2000) which are based on the clumping-corrected
mass loss rates of Nugis et al. (1998). Specifically for the WN stars,
we used the relationship
\[ \log \dot M = -13.60 + 1.63 \log \frac{L}{L_\odot} +2.22 \log Y \]
which agrees well with results of line-blanketed analyses (Crowther
2001). For WC stars, Nugis \& Lamers (2000) find that their mass loss
rates do show an increase with luminosity but the scatter is
large. Since we have chosen constant values of $L$ and $Y$ for all WC
subtypes we used the WC linear regression relation of Nugis \& Lamers
(2000) to give a fixed mass loss rate of $\log \dot M=-4.72$ at solar
metallicity. To assign terminal velocities, we determined approximate
spectral types and used the $v_\infty$--spectral type calibrations of
Prinja et al. (1990).

It is not known whether the wind parameters of W-R stars scale with
metallicity because the driving mechanism of their winds has not been
conclusively determined, and their atmospheres are chemically enriched
in the products of the CNO-cycle (WNs) and He-burning (WCs).  
In the past, it has usually been assumed that there is no scaling with 
metallicity. 

Observationally, the parameters of Galactic and LMC WN stars do not
show a clear metallicity dependence (Hamann \& Koesterke 2000)
although the factor of $\sim$ two change in metallicity may not be sufficient
to reveal significant variations within the observational and
modelling uncertainties. Crowther (2000) analysed one SMC WN star and
again found no differences relative to comparable stars in the LMC and
Galaxy, although the spread in properties was large. More recently,
Crowther et al. (2002a) have studied a sample of LMC WC4 stars and
derived wind densities  which were $\sim$0.2 dex lower
than Galactic WC stars analysed in the same manner (Dessart et
al. 2000). Crowther et al. concluded that a metallicity dependence for
W-R stars naturally explains the well known difference between the
Galactic and LMC WC populations, since the classification diagnostic
C\three\ $\lambda5696$ is extremely mass-loss dependent.

Recent theoretical work also indicates that the strength of W-R winds
should be metallicity dependent. Although the precise wind driving
mechanism is still unsolved, it is generally believed that radiation
pressure with multiple scattering is responsible. The reduction in
mass loss rates through clumping has alleviated the momentum problem
but the momentum transfer efficiencies still lie in the range 2--20
(Nugis \& Lamers 2000). Heavy metals are critical for initiating the
outflow, even for WC stars, since the high ions of C and O do not
possess enough driving lines in the correct energy range (Crowther et
al. 2002a). Therefore we have adopted the same power law exponent
scaling as used previously for the O star models (Sect.~\ref{Ogrid}).
The weak metallicity effect identified by Crowther et al. (2002a) is
consistent with the scaling in our approach.

\begin{figure}
\psfig{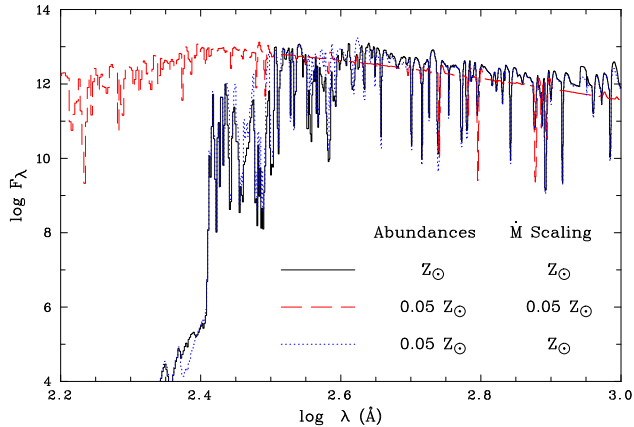}
\caption{The flux distribution of model WC\#10 ($T_*=100\,000$\,K) for three 
different assumptions about the abundances and mass loss scaling: (a)
abundances and mass loss rate scaled to solar (black, solid); (b) abundances
and mass loss rate scaled to $0.05$\,Z$_\odot$ (red, dashed); and (c)
$0.05$\,Z$_\odot$ abundances but mass loss rate scaled to solar (blue,
dotted). Only model (b) with wind parameters scaled to $0.05$\,Z$_\odot$
has a significant flux below the He$^+$ edge at 228\,\AA\ because of
the reduced wind density. The flux is in units of
ergs\,cm$^{-2}$\,s$^{-1}$\,\AA$^{-1}$ measured at the stellar
surface.}
\label{mdot}
\end{figure}

Observations show that nebular He\two\ $\lambda4686$ emission is
spatially associated with hot W-R stars in low metallicity
environments (Garnett et al. 1991; Schaerer, Contini \& Pinado 1999).
Schmutz et al. (1992; hereafter SLG92) discuss the fact that the
energy output above 54.4\,eV depends on the mass loss rate. In a W-R
star with a strong wind, He$^{++}$ recombines at some point, whereas
there is no recombination in a weak wind, and flux is emitted above
54.4\,eV. In Fig.~\ref{mdot}, we show the flux distribution of a WC
model with $T_*=100\,000$\,K for three different assumptions about the
line blanketing and wind density. It is apparent that for a solar
(i.e. unscaled) mass loss rate, no ionizing photons escape below
228\,\AA\ at $Z=0.05$ and $Z=$Z$_\odot$.  Indeed, the emergent spectra
are very similar, demonstrating that the reduced line blanketing at
$Z=0.05$\,Z$_\odot$ has a small effect. In contrast, the model which has
wind parameters scaled to $Z=0.05$\,Z$_\odot$ has a significant ionizing
flux below the He$^+$ edge. We thus find that the wind density
controls the transparency of the wind below 228\,\AA\ at a given
stellar temperature and the effect of line blanketing is
negligible. The photon luminosities of the models are given in Tables
3 and 4 in the form of $\log Q_0, \log Q_1$ and $\log Q_2$. It should
be noted that no models with 2\,Z$_\odot$ abundances have significant
$Q_2$ fluxes.

The mass loss rates in Tables 3 and 4 are corrected for clumping since
they are derived from the Nugis \& Lamers (2000) relationship. For
consistency, we have introduced clumping into the {\sc cmfgen} model
atmosphere calculations by adopting a volume filling factor $f$ of 0.1
(Hillier \& Miller 1999). This has no effect on the emergent flux but
does make a difference to the shape of the W-R line profiles. It is
possible to recover the unclumped mass loss rates by multiplying the
values in Tables 3 and 4 by $1/f^{1/2}$, i.e. $\sqrt{10}$. For these
calculations, we assume a standard velocity law of $v(r) =
v_\infty(1-R/r)^\beta$ with $\beta=1$, and a Doppler line width of
200 km\,s$^{-1}$. 

Since the wind density is such a critical parameter, our use of the
Nugis \& Lamers (2000) mass-loss-luminosity calibration deserves
special mention. As discussed above, this reproduces the average
mass-loss rates for strong-lined Galactic W-R stars rather well. The
weak-lined W-R stars, however, fall up to $\sim$0.5 dex below this
calibration. Some pathological stars have even weaker winds and are
not accounted for in our calculations. Examples are HD\,104994 (WN3)
which Crowther, Smith \& Hillier (1995) predicted to emit strongly in
the He$^{+}$ continuum, whilst nebular He\,{\sc ii} $\lambda$4686 is
observed in the H\,{\sc ii} region G\,2.4+1.4 which is associated with
Sand~4 (WO, Esteban et al. 1992). LMC W-R stars known to emit strongly
below $\lambda$228\,\AA\ also exist, such as Brey~2 (WN2, Garnett \&
Chu 1994).  Fortunately, such W-R stars are relatively few in number,
so it is probable that neglecting their contribution is acceptable for
the large massive star population in a starburst.
\begin{figure}
\psfig{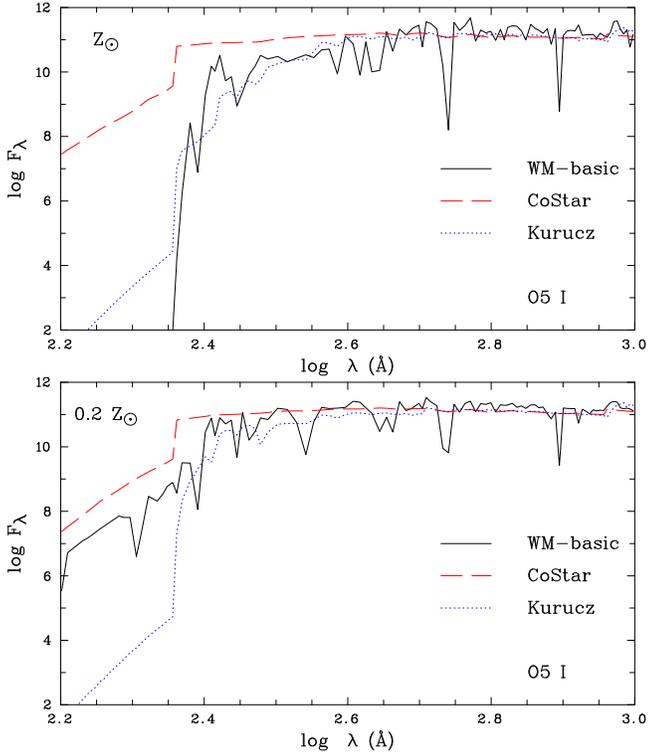}
\caption{Comparison of the emergent fluxes for an early O supergiant 
from a WM-basic (black, solid), a CoStar (red, dashed), and a Kurucz
model (blue, dotted) at solar metallicity and $0.2$\,Z$_\odot$.  The model
details for $Z_\odot$ are: WM-basic OB\#14 $T_{{\rm
eff}}=42.6$\,kK; $\log g=3.67$; $R_*=19.7$\,R$_\odot$; $\log \dot M
=-4.92$; $v_\infty=2300$\,km\,s$^{-1}$; CoStar \#E3 $T_{{\rm
eff}}=42.6$\,kK; $\log g=3.71$; $R_*=20.0$\,R$_\odot$; $\log
\dot M =-4.88$; $v_\infty=2538$\,km\,s$^{-1}$; and Kurucz $T_{{\rm
eff}}=42.5$\,kK; $\log g=5.0$. The model parameters for $0.2$\,Z$_\odot$
are scaled as discussed in the text.}
\label{O5I}
\end{figure}
\begin{figure}
\psfig{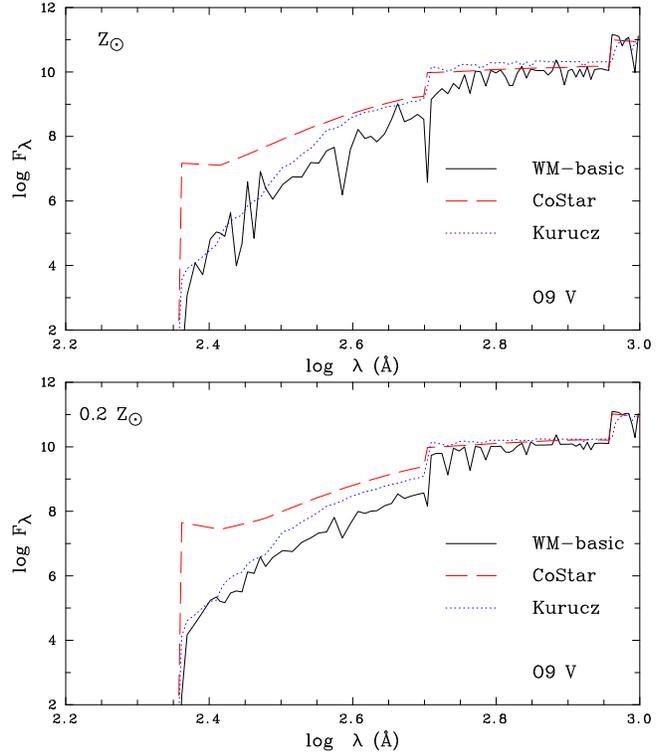}
\caption{Comparison of the emergent fluxes for a late O dwarf
from a WM-basic (black, solid), a CoStar (red, dashed), and a Kurucz
model (blue, dotted) at solar metallicity and $0.2$\,Z$_\odot$.  The model
details for $Z_\odot$ are: WM-basic OB\#7 $T_{{\rm
eff}}=32.3$\,kK; $\log g=4.00$; $R_*=10.1$\,R$_\odot$; $\log \dot M
=-6.74$; $v_\infty=1500$\,km\,s$^{-1}$; CoStar \#A2 $T_{{\rm
eff}}=33.3$\,kK; $\log g=4.01$; $R_*= 7.3$\,R$_\odot$; $\log
\dot M =-6.86$; $v_\infty=2544$\,km\,s$^{-1}$; and Kurucz $T_{{\rm
eff}}=32.0$\,kK; $\log g=4.0$. The model parameters for $0.2$\,Z$_\odot$
are scaled as discussed in the text.}
\label{O9V}
\end{figure}
 
\section{Model Atmosphere Comparisons}\label{comp}
\subsection{O star models}\label{Ocomp}
The first expanding, line-blanketed, non-LTE grid of O star model
atmospheres was introduced by Schaerer \& de Koter (1997; hereafter
SK97). These combined stellar structure and atmosphere or `CoStar'
models are used extensively in single star H\two\ region analyses
(e.g. Oey et al. 2000) and spectral synthesis studies (e.g. Stasi\'
nska et al. 2001). The CoStar grid consists of 27 models covering the
main sequence evolution of O3 to early B spectral types at
metallicities of solar and 0.2\,Z$_\odot$. The mass loss rates are
taken from Meynet et al. (1994) and are scaled with metallicity using
an exponent of 0.5. To compare the WM-basic with the CoStar models, we
have chosen a hot (T$_{{\rm eff}}=43$\,kK; OB\#14) early O supergiant
model with a strong wind, and a cool (T$_{{\rm eff}}=33$\,kK; OB\#7)
late O dwarf model with a weak wind. In Figs.~\ref{O5I} and \ref{O9V}
we show the comparisons at solar and 0.2\,Z$_\odot$ and also plot
Kurucz (1992) LTE plane parallel models for reference. The details of
the model parameters are given in the figure captions.

There are clearly significant differences in the emergent fluxes from
the WM-basic and CoStar models. Considering first the O supergiant
case, the WM-basic model has no flux below the He$^+$ edge at
Z$_\odot$ in contrast to the CoStar model whereas at 0.2\,Z$_\odot$,
both models have flux below 228\,\AA. We find again that the wind
density is the critical parameter in determining the transparency
below 228\,\AA\ for the WM-basic models; a model with an un-scaled
(i.e. solar) mass loss rate and 0.2\,Z$_\odot$ abundances has no flux
below this wavelength (cf. Section~\ref{W-Rgrid}). The reason for the
difference between the CoStar and WM-basic models at $Z_\odot$ is less
clear because both models have the same wind density. Crowther et
al. (1999) explore the differences in the ionizing fluxes predicted by
{\sc cmfgen} and the {\sc isa}-wind code of de Koter, Heap \& Hubeny
(1997) which is used for the CoStar atmosphere calculations.  They
suggest that the most likely reason for the difference is that {\sc
isa}-wind overestimates the far-UV flux because photon line-absorption
and subsequent re-emission at longer wavelengths is not taken into
account.

For the late O dwarf case, we find that the WM-basic model has a lower
flux in the He$^0$ continuum compared to both the CoStar and Kurucz
models.  It is closer to the Kurucz model, as expected for cool dwarf
stars with low mass loss rates. The lower flux compared to the Kurucz
model is probably due to non-LTE effects producing deeper line cores
in the blocking lines (Sellmaier et al. 1996).  SK97 investigated the
excess flux in their dwarf models compared to plane parallel non-LTE
models. They find that the temperature is too high in the region where
the He$^0$ continuum is formed and caution against the use of these
models.

\subsection{W-R models}\label{W-Rcomp}
As discussed in Section~\ref{W-Rgrid}, the W-R grid is based on
current best estimates of wind parameters, temperatures, luminosities
and chemical compositions. These parameters differ substantially from
the pure helium W-R model grid of Schmutz et al. (1992) which is
implemented in the Starburst99 evolutionary synthesis code.  
The SLG92 grid is defined by three parameters: the `transformed
radius' $R_{\rm t}$ (a measure of the inverse wind density) given by
\[ R_{\rm t} = R_* \left( \frac{10^{-4}}{dM/dt} \ \ 
\frac{v_\infty}{2500}\right)^{2/3},\]
the stellar or core effective temperature $T_*$, and the velocity law
exponent ($\beta=1$ or 2).  
\begin{figure}
\psfig{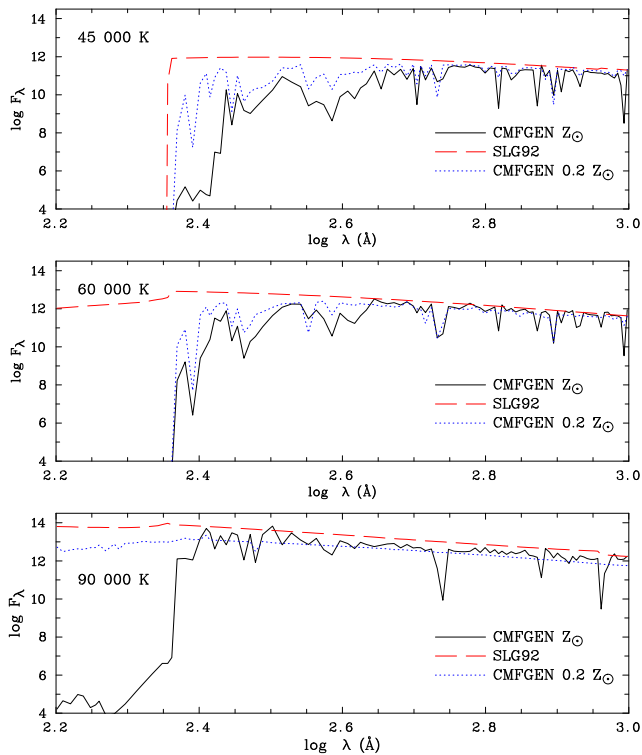}
\caption{
The emergent fluxes for {\sc cmfgen} WN models with $T_* = 45, 60$ and
90\,kK for $Z= $Z$_\odot$ (black, solid) and 0.2\,Z$_\odot$ (blue,
dotted). For comparison, the SLG92 models are also plotted (red,
dashed) that would be chosen in Starburst99 for the same evolutionary
point at 4\,Myr in the HRD but corresponding to the hydrostatic core Meynet
et al. temperatures of 56.2, 82.2 and 130.1\,kK.  The transformed
radii $R_{\rm t}$ of these three models are 17.9, 8.4 and
1.7\,R$_\odot$.}
\label{WNcomp}
\end{figure}
In Starburst99, SLG92 W-R atmospheres are connected to the
evolutionary tracks by interpolating $T_*$ and $R_{\rm t}$ to match
the hydrostatic effective temperature $T_{\rm {hyd}}$ tabulated in
Meynet et al. (1994) and the $\dot M$ and $v_\infty$ values of
Leitherer et al. (1992) as a function of W-R type. As discussed in the
next section, the interfacing of W-R atmospheres to evolutionary
tracks is difficult because of the optically thick nature of the W-R winds. We have
chosen to implement the {\sc cmfgen} W-R atmospheres by matching
$T_*$ to $0.6 T_{\rm {hyd}} + 0.4 T_{2/3}$ where $T_{2/3}$ represents
the corrected temperature tabulated by Meynet et al. (1994).  To make
a meaningful comparison between the SLG92 and the new W-R grids, we
compare models that would be selected in Starburst99 and our updated
version to represent a W-R star at a given evolutionary point in the
H-R diagram.
\begin{figure}
\psfig{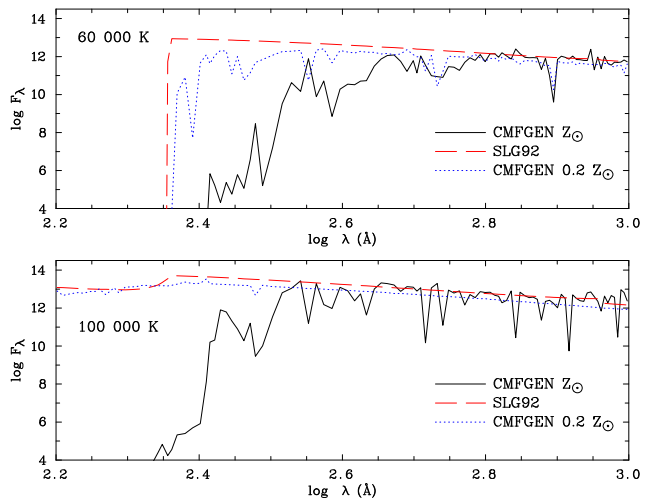}
\caption{
The emergent fluxes for {\sc cmfgen} WC models with $T_* = 60$ and
100\,kK for $Z= $Z$_\odot$ (black, solid) and 0.2\,Z$_\odot$ (blue,
dotted). For comparison, the SLG92 models are also plotted (red,
dashed) that would be chosen in Starburst99 for the same evolutionary
point at 5\,Myr in the HRD but corresponding to the hydrostatic Meynet
et al. temperatures of 78.2 and 126.1\,kK.  The transformed radii
$R_{\rm t}$ of these two models are 2.7 and 1.6\,R$_\odot$.}
\label{WCcomp}
\end{figure}

In Fig.~\ref{WNcomp}, we show three {\sc cmfgen} WN models with $T_*=
45\,000, 60\,000$ and 90\,000\,K for metallicities of solar and
0.2\,Z$_\odot$. For comparison, we plot the equivalent SLG92 models
that would be chosen in Starburst99 at an age of 4\,Myr corresponding
to $T_{\rm{hyd}}= 56\,200, 82\,200$ and 130\,100\,K. Significant
differences are seen because of the different wind densities, our
inclusion of line blanketing, and the lower temperatures we adopt to
match the W-R atmospheres to the evolutionary tracks.  For the
45\,000\,K model, the emergent flux below 504\,\AA\ is much lower for
the {\sc cmfgen} model at $Z_\odot$ compared to SLG92 because of line
blanketing, as shown by the increased flux in the $0.2$\,Z$_\odot$
model. The 60\,000\,K {\sc cmfgen} model has no flux below 228\,\AA\
in contrast to the higher temperature SLG92 model, and the flux below
504\,\AA\ is again reduced. The large difference in the emergent flux
below 228\,\AA\ seen in Fig.~\ref{WNcomp} at 90\,000\,K at $Z_\odot$
between the {\sc cmfgen} and SLG92 model is due to the wind
density effect discussed in Section~\ref{W-Rgrid}, since the
$0.2$\,Z$_\odot$ {\sc cmfgen} model has emergent flux below 228\,\AA.

In Fig.~\ref{WCcomp}, we plot two {\sc cmfgen} WC models
with temperatures of $T_*=60\,000$ and 100\,000\,K and compare them to
the models that would be chosen in
Starburst99 at an age of 5\,Myr corresponding to $T_{\rm{hyd}} = 78\,200$ and
126\,100\,K.  The differences seen for the cooler case are similar to
the 45\,000\,K WN model comparison in that the line blanketing at
$Z_\odot$ produces a significant reduction in the flux in the He$^0$
continuum. The differences seen at 100\,000\,K can be explained by the
wind density determining the flux below 228\,\AA.

\begin{figure}
\psfig{file=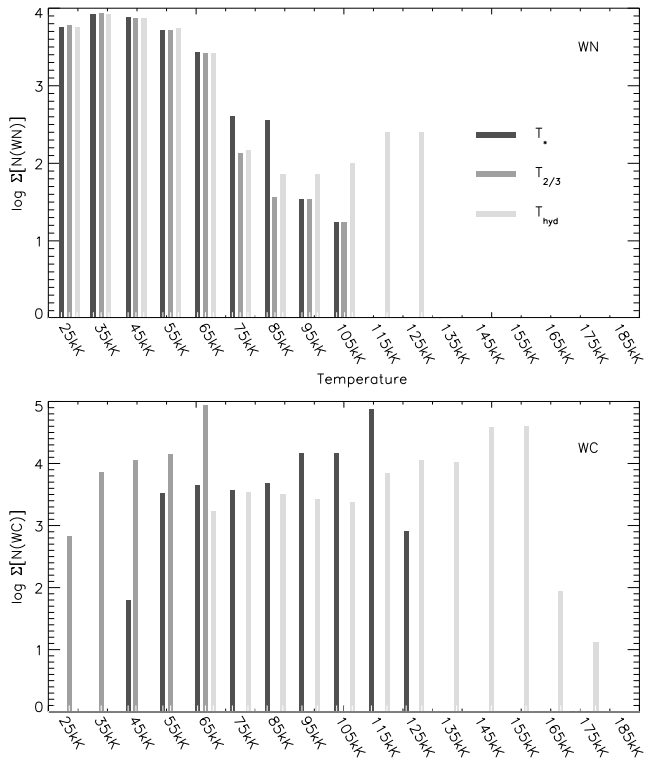,width=8.4cm,bbllx=89pt,bblly=152pt,bburx=542pt,bbury=691pt}
\caption{Histogram plots of the temperature distribution for the WN and WC phases
summed over the lifetime of the W-R phase as given by Starburst99 with
the Meynet et al. (1994) evolutionary tracks at solar metallicity for
an instantaneous burst with $M=10^6$\,M$_\odot$. For each W-R
type, we show the temperature distributions corresponding to the
hydrostatic temperature $T_{\rm {hyd}}$ (light grey), the corrected
hydrostatic temperature $T_{2/3}$ (dark grey) of Meynet et al. (1994),
and the adopted $T_* = 0.6 T_{\rm {hyd}} + 0.4 T_{2/3}$ (black). The temperature
bin sizes are 10 kK.}
\label{temp}
\end{figure}

\section{Integration into Starburst99}\label{sb99}
Leitherer et al. (1999) present the evolutionary synthesis code
Starburst99 which predicts the observable properties of galaxies
undergoing active star formation, and is tailored to the analysis of
massive star populations. The code is an improved version of that
previously published by Leitherer \& Heckman (1995). It is based on
the evolutionary tracks of Meynet et al. (1994) for enhanced mass loss
and the model atmosphere grid compiled by Lejeune et al. (1997),
supplemented by the pure helium W-R atmospheres of Schmutz et
al. (1992).

To integrate the new O star grid, we have simply replaced the Lejeune
et al.  (1997) LTE models by the WM-basic models for effective
temperatures above 25\,000\,K, re-mapped to 1221 points over the
wavelength range $\log
\lambda=1.96$--6.20\,\AA, as required by the code.  To ensure a smooth
transition to the LTE models, we adopt the method of SV98: the
WM-basic models are restricted to the $T_{{\rm eff}}$--$\log g$ domain
defined by $\log g \ge 2.2$ and $\log g < 5.71 \times \log T_{{\rm
eff}} - 21.95$. The ionizing fluxes of giants were determined by
averaging the supergiant and dwarf models, and renormalising to
conserve the flux. Tests using real giant models show that this method
is sufficiently accurate. We adopt the same method used by Leitherer
et al. (1999) in choosing the best model atmosphere to represent an
evolutionary point viz. selection using the nearest fit in $\log g$
and $T_{\rm {eff}}$ and then interpolation, rescaling for $T_{{\rm
eff}}$ and $R$.

The new W-R grid is based on current best estimates of W-R
temperatures and these are considerably lower than those of SLG92 and
the evolutionary model temperatures. The problem of interfacing W-R
atmospheres with evolutionary tracks is an unresolved issue which
SLG92 discuss in some detail. It exists because the temperature given
by evolutionary models $T_{\rm {hyd}}$ refers to the hydrostatic
core. Since W-R winds are optically thick, the observed radiation
emerges at larger radii, and the temperature derived from
observational analyses cannot be related to $T_{\rm {hyd}}$ without
inward extrapolation. To overcome this, Maeder (1990) and Meynet et
al. (1994) correct the hydrostatic $T_{\rm {hyd}}$ value by assuming a
velocity law to give the radius corresponding to an optical depth of
2/3 and hence a lower temperature $T_{2/3}$. This does not, however,
solve the problem because the corresponding $T_{2/3}$ temperatures
derived from W-R model atmospheres analyses are generally close to
30\,000\,K irrespective of W-R subtype. As discussed by SLG92, W-R
model atmospheres are best characterized by the core temperature $T_*$
which refers to the radius where the optical depth is $\approx$ 10
(corresponding to a thermalization depth of unity for the continuum
opacity which is dominated by electron scattering).

The problem is therefore how to assign a W-R model atmosphere
characterized by $T_*$ to an evolutionary model defined by two
temperatures: $T_{\rm {hyd}}$ and $T_{2/3}$, with the former being too
hot and the latter too cool. In Starburst99, an SLG92 W-R atmosphere
is assigned by interpolating $T_*$ and $R_{\rm t}$ to match $T_{\rm
{hyd}}$ of the Meynet et al. (1994) evolutionary models.  Since the
$T_{\rm {hyd}}$ values of Meynet et al. are higher than the new W-R
grid values, we have assigned the W-R models to the tracks by matching
$T_*$ to $0.6 T_{\rm {hyd}} + 0.4 T_{2/3}$.  We arrived at this
formalism by studying the distribution of temperatures in the W-R
phase as given by Starburst99 for solar metallicity. We simply binned
the number of WN and WC stars into 10\,kK temperature bins and summed
over the lifetime of the W-R phase. The resulting WN and WC
temperature distributions are shown in Fig.~\ref{temp}.

For the WN stars, the temperature distribution for $T_{\rm {hyd}}$
shows one broad peak at $\sim 40\,000$\,K and a second peak near
120\,000\,K. The distribution for the WC stars is flatter and peaks at
$\sim 150\,000$\,K.  We first tried a simple average of the two Meynet
et al.  temperatures but this shifted the peak of the WN temperature
distribution to a cool $\sim 30\,000$\,K. We then adjusted the bias
between the two temperatures by trial and error and found that $0.6
T_{\rm {hyd}} + 0.4 T_{2/3}$ gave a single temperature distribution
for the WNs peaking at 40\,000\,K and extending to 100\,000\,K. The
corresponding WC temperature distribution, as shown in
Fig.~\ref{temp}, peaks at 110\,000\,K. Ideally, the next step would be
to compare the adopted $T_*$ distribution with observed W-R subtype
vs. temperature distributions. Unfortunately there is insufficient
data to do such a comparison. Nevertheless, our adopted $T_*$ values
for WN and WC stars span the range occupied by Galactic and LMC W-R
stars, as derived by recent analyses (e.g. Hillier \& Miller 1998;
Herald et al. 2001; Crowther et al. 2002a).  We also note that the WN
and WC $T_*$ peaks correspond to late WN and early WC stars. Studies of W-R
galaxies (e.g. Guseva, Izotov \& Thuan 2000) show that the composite
W-R feature at $\sim 4680$\,\AA\ is usually dominated by these types
of W-R stars.

Finally, we have adopted the same switching technique employed by
Leitherer et al. (1999): W-R atmospheres are used when the surface
temperature of the Meynet et al. (1994) tracks exceeds 25\,000\,K and
the surface hydrogen content is less than 0.4. In addition, a WN or WC
model is chosen depending on whether C/N$<10$ (WN) or $>10$ (WC).
\begin{figure*}
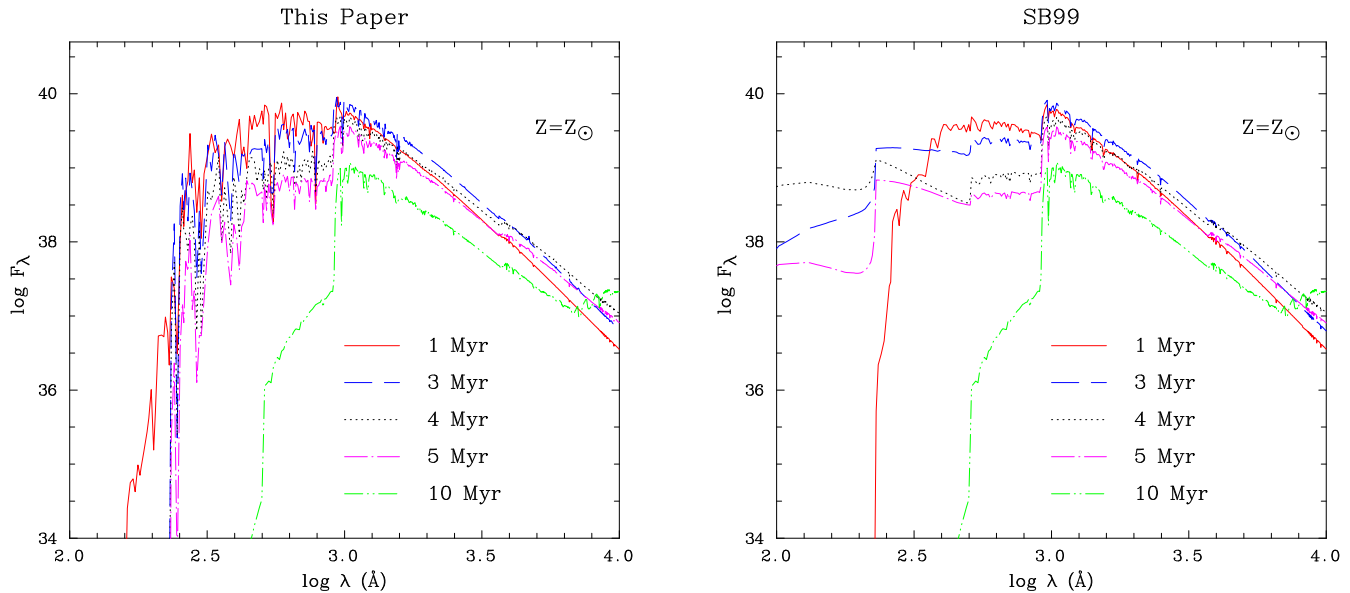

\begin{minipage}{8.4cm}
\psfig{file=fig_sed_us_col.eps,width=8.3cm,bbllx=77pt,bblly=267pt,bburx=439pt,bbury=610pt}
\end{minipage}
\hspace*{0.9cm}
\begin{minipage}{8.4cm}
\psfig{file=fig_sed_sb99_col.eps,width=8.3cm,bbllx=77pt,bblly=267pt,bburx=439pt,bbury=610pt}
\end{minipage}
\caption{Spectral energy distributions of the new models (left-hand side)
compared with Starburst99 (right-hand side) at time intervals of 1, 3,
4, 5 and 10\,Myr for an instantaneous burst at $Z_\odot$.}
\label{sed_inst}
\end{figure*}
\section{Evolutionary Synthesis Comparisons}\label{evolcomp}
In Section~\ref{comp}, we compared the new O and W-R model atmospheres
with the grids of Schaerer \& de Koter (1997) and Schmutz et
al. (1992).  We find that there are significant differences below
228\,\AA\ at $Z_\odot$ (and higher) for the hot O supergiant and W-R
models.  We can thus expect large differences in the output ionizing
fluxes of young massive star-forming regions, especially during the W-R
phase. We compare the emergent flux distributions obtained with the
new models integrated into Starburst99 to:
\begin{enumerate}
\item
the standard version of Starburst99 (which we will refer to as SB99)
which uses the W-R atmospheres of Schmutz et al. (1992) for stars with
strong mass loss and the LTE models from the compilation of Lejeune et
al. (1997) for all other stars.
\item
a modified version of (i) with the CoStar models of Schaerer \& de
Koter (1997) replacing the LTE models for the main sequence evolution of
massive stars. We have implemented these models as described in Schaerer
\& Vacca (1998) and use the solar metallicity grid for solar and above,
and the $0.2$\,Z$_\odot$ grid for $Z \le 0.4$\,Z$_\odot$. We refer to
these models as SV98 and have checked that the output is the same as
the actual SV98 models for the hot O star phase. There are small
differences in the W-R phase resulting from the different methods of
interpolation used in the SB99 and SV98 codes.
\end{enumerate}
We consider two types of star formation: an instantaneous burst and
continuous star formation.  We adopt a standard initial mass function
(IMF) with a single Salpeter power law slope of 2.35 and lower and
upper mass cut-offs of 1 and 100\,M$_\odot$. We use the enhanced mass
loss tracks of Meynet et al. (1994) and calculate models for five
metallicities: 0.05, 0.2, 0.4, 1 and 2\,$Z_\odot$. We do not include
the contribution of the nebular continuum. 
\begin{figure*}
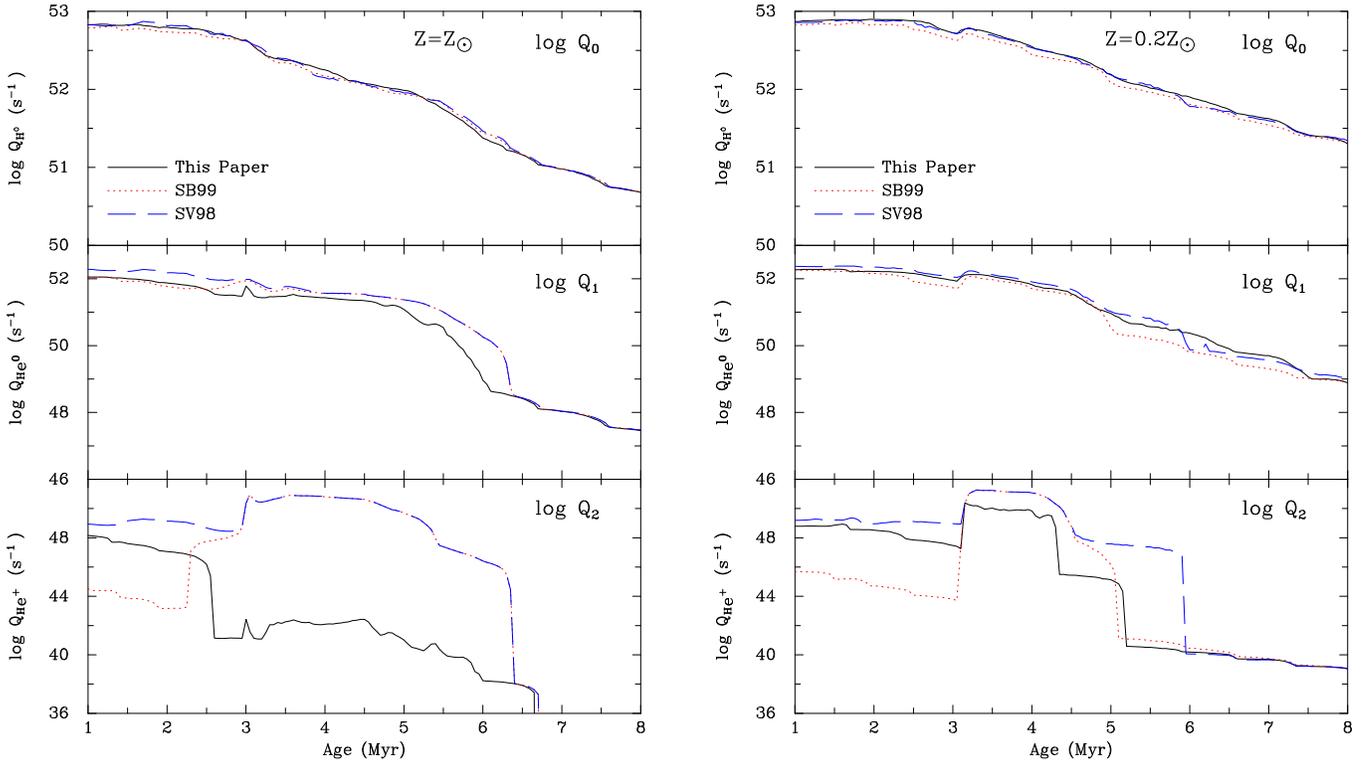

\begin{minipage}{8.4cm}
\psfig{file=fig_qsol_col.eps,width=8.3cm,bbllx=77pt,bblly=207pt,bburx=437pt,bbury=643pt}
\end{minipage}
\hspace*{0.9cm}
\begin{minipage}{8.4cm}
\psfig{file=fig_qsmc_col.eps,width=8.3cm,bbllx=77pt,bblly=207pt,bburx=437pt,bbury=643pt}
\end{minipage}
\caption{
The evolution of the photon luminosity for ages of 1--8\,Myr in the
ionizing continua of hydrogen ($\log Q_0$), He\one\ ($\log Q_1$) and
He\two\ ($\log Q_2$) at 0.2 and $Z_\odot$ for an instantaneous burst
at time steps of 0.1\,Myr. The new models (black, solid) are compared to the
ionizing fluxes of: Starburst99 (SB99, red, dotted) using the stellar
library of Lejeune et al. (1997) and the Schmutz et al. (1992) W-R
models; the Schaerer \& Vacca (1998) models (SV98, blue, dashed) using
CoStar and Schmutz et al. (1992) W-R atmospheres.}
\label{qs_inst}
\end{figure*}
\begin{figure*}
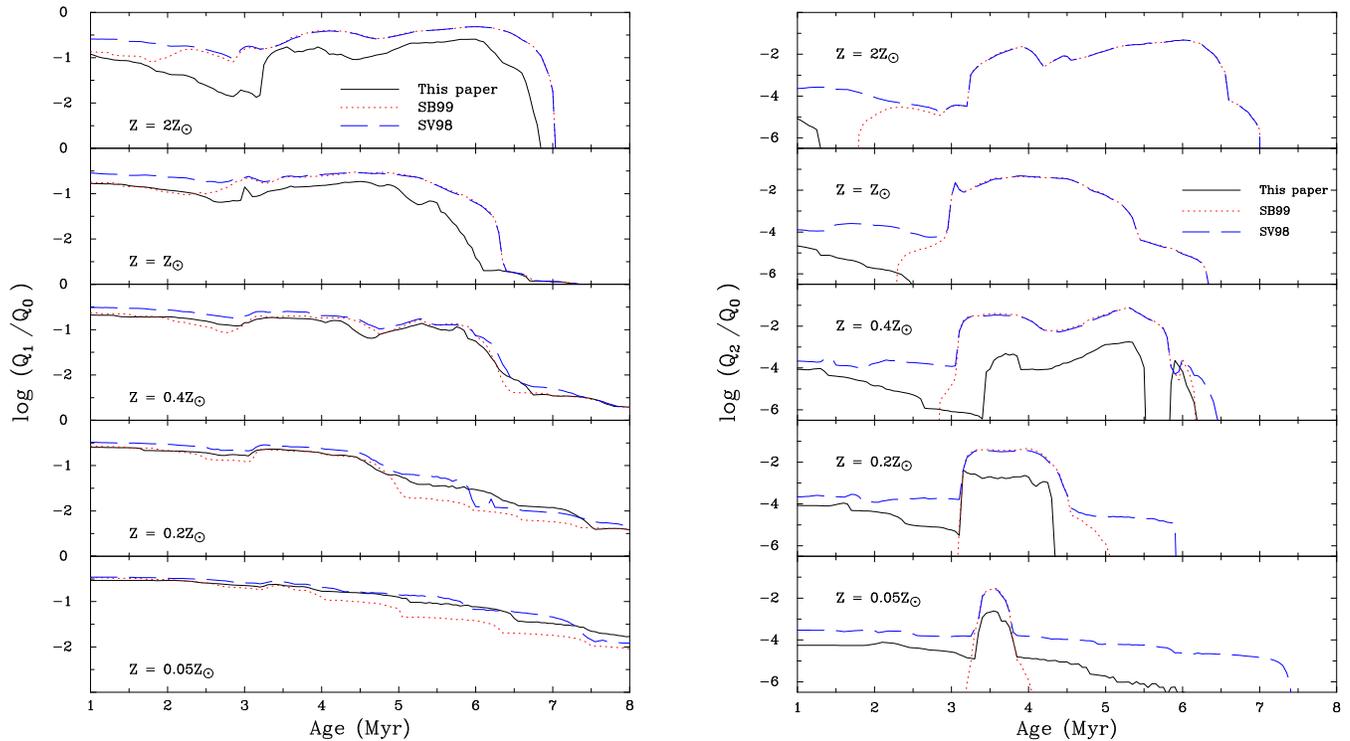

\begin{minipage}{8.4cm}
\psfig{file=fig_q1q0_col.eps,width=8.3cm,bbllx=64pt,bblly=177pt,bburx=433pt,bbury=616pt}
\end{minipage}
\hspace*{0.9cm}
\begin{minipage}{8.4cm}
\psfig{file=fig_q2q0_col.eps,width=8.3cm,bbllx=64pt,bblly=177pt,bburx=433pt,bbury=616pt}
\end{minipage}
\caption{The evolution of the hardness of the ionizing
spectra shown by the ratios of $\log (Q_1/Q_0)$ (lefthand side) and
$\log (Q_2/Q_0)$ (righthand side) at all five metallicities for an
instantaneous burst and ages of 1--8\,Myr. The details of the models
are given in Fig.~\ref{qs_inst}.}
\label{qrs_inst}
\end{figure*}

\subsection{Instantaneous burst}\label{inst}
We adopt a standard instantaneous burst model with a total mass of
$10^6$\,M$_\odot$. In Fig.~\ref{sed_inst}, we show the spectral energy
distributions obtained with the new model atmospheres compared with
SB99 at time intervals of 1, 3, 4, 5 and 10\,Myr for $Z_\odot$.  The
most dramatic differences occur below the He$^+$ edge at 228\,\AA\
during the W-R phase at 3--5\,Myr. The new line-blanketed models have
negligible flux below 228\,\AA\ at $Z_\odot$ in contrast to the very
hard ionizing fluxes of the SLG92 atmospheres.  The detailed
differences between the new models and previous work, and their
dependence on metallicity are best seen by considering the ionizing
fluxes and their ratios.  In Fig.~\ref{qs_inst}, we show the evolution
of the photon luminosity from 1--8\,Myr in the ionizing continua of
H\one\ ($\log Q_0$), He\one\ ($\log Q_1$) and He\two\ ($\log Q_2$) for
0.2 and $Z_\odot$. The evolution of the hardness of the ionizing
spectra $\log (Q_1/Q_0)$ and $\log (Q_2/Q_0)$ for all five
metallicities is shown in Fig.~\ref{qrs_inst}.  The variation with
metallicity is primarily due to the changing W-R/O ratio.  The
relative number of W-R stars formed and the duration of the W-R phase
increases with metallicity because the lower mass limit for their
formation decreases (Meynet 1995). The W-R/O ratio changes from 0.025
to 0.36 between $Z=0.05$--2\,Z$_\odot$ at 3\,Myr (SV98).
\begin{figure*}
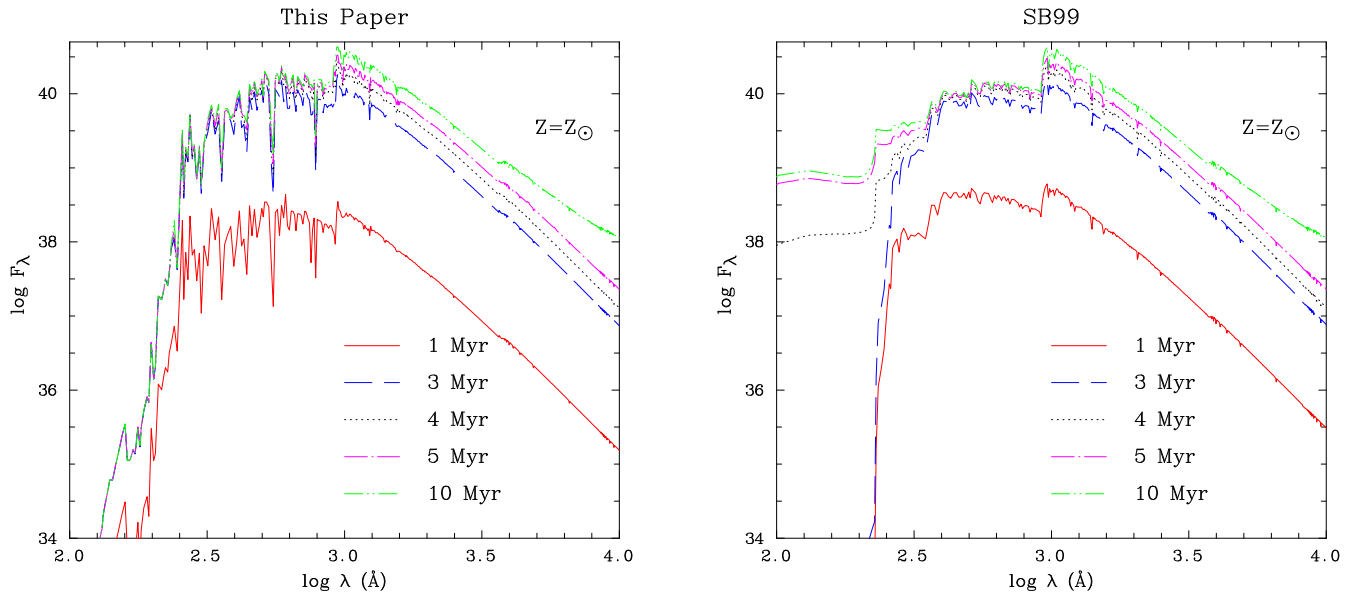

\begin{minipage}{8.4cm}
\psfig{file=fig_sed_us_cns_col.eps,width=8.3cm,bbllx=77pt,bblly=267pt,bburx=439pt,bbury=610pt}
\end{minipage}
\hspace*{0.9cm}
\begin{minipage}{8.4cm}
\psfig{file=fig_sed_sb99_cns_col.eps,width=8.3cm,bbllx=77pt,bblly=267pt,bburx=439pt,bbury=610pt}
\end{minipage}
\caption{Spectral energy distributions of the new models (left-hand side)
compared with Starburst99 (right-hand side) at time intervals of 1, 3,
4, 5 and 10\,Myr for continuous star formation at $Z_\odot$.
}
\label{sed_cns}
\end{figure*}

The ionizing flux in the Lyman continuum $Q_0$ is determined mainly by
hot main sequence stars. Fig.~\ref{qs_inst} shows that $Q_0$ is very
similar to SV98 although the SB99 values are slightly lower because
the Kurucz LTE models for hot O stars have smaller radii than the
non-LTE O star models.  The new models have lower ionizing fluxes
below 504\,\AA\ at ages of less than $\sim$ 7\,Myr for metallicities
$\ge 0.4$\,Z$_\odot$. This is because W-R stars contribute an
important fraction of the He\one\ ionizing flux and our new W-R models
are line-blanketed compared to the SLG92 grid used for the SB99 and
SV98 models. The new O star models also have a lower ionizing flux in
the He\one\ continuum (Sect.~\ref{Ocomp}).  The dependence of $\log
(Q_1/Q_0)$ on metallicity is demonstrated in Fig.~\ref{qrs_inst}. 
At 5\,Myr, the ratio $Q_1/Q_0$ is softer by a factor of 2 at
2\,Z$_\odot$ compared to SB99 and SV98 because of the effect of the
W-R line-blanketed atmospheres, whereas at $Z=0.05$\,Z$_\odot$, the
models agree with SV98 because at this low metallicity, the main
contributors to $Q_1$ are O supergiants because of the low W-R/O
ratio. In general, at sub-solar metallicities for $Q_1/Q_0$, the new
models agree more closely with SV98 than SB99.

The most pronounced revisions are seen in $Q_2$ during the W-R phase at
3--6\,Myr in Fig.~\ref{qs_inst} for Z$_\odot$. The number of He\two\
ionizing photons emitted is now negligible compared to SV98 and SB99
because of the W-R wind density effect (Sect.~\ref{W-Rgrid}) and the
conspicuous bump in the SV98 and SB99 models disappears. 
At metallicities below solar, $Q_2/Q_0$ (Fig.~\ref{qrs_inst}) is
softer in the W-R phase compared to SV98 and SB99 by, for example, a
factor of $\sim 25$ at 0.4\,Z$_\odot$. The W-R subtypes can be
identified since we have used different models for the WN and WC
phases.  At 0.4\,Z$_\odot$, the first W-R stars to form at 3.5\,Myr are
the late WN stars, the descendants of the most massive O stars. These
stars then evolve to WC stars, and finally, near the end of the W-R
phase, the O stars near the minimum mass limit for W-R formation,
become WN stars. At $Z=0.05$\,Z$_\odot$, the W-R phase is short-lived
because only the most massive O stars evolve into W-R stars.  At all
metallicities, the new O star models are softer compared to SV98
because the far-UV flux in the CoStar atmospheres is probably
overestimated (see discussion in Sect.~\ref{Ocomp}). 

Overall, at metallicities of solar and higher, we find that the
He\two\ ionizing flux is negligible because of the high wind densities
and significant line-blanketing. He\two\ ionizing photons are produced
during the W-R phase at metallicities below solar but $Q_2/Q_0$ is much
softer by a factor of $\sim 20$ compared to the SLG92 W-R models.  In
Section~\ref{discuss}, we discuss the presence of nebular He\two\
$\lambda4686$ as an indicator of a significant W-R population and its
strength as a function of metallicity.
\begin{figure*}
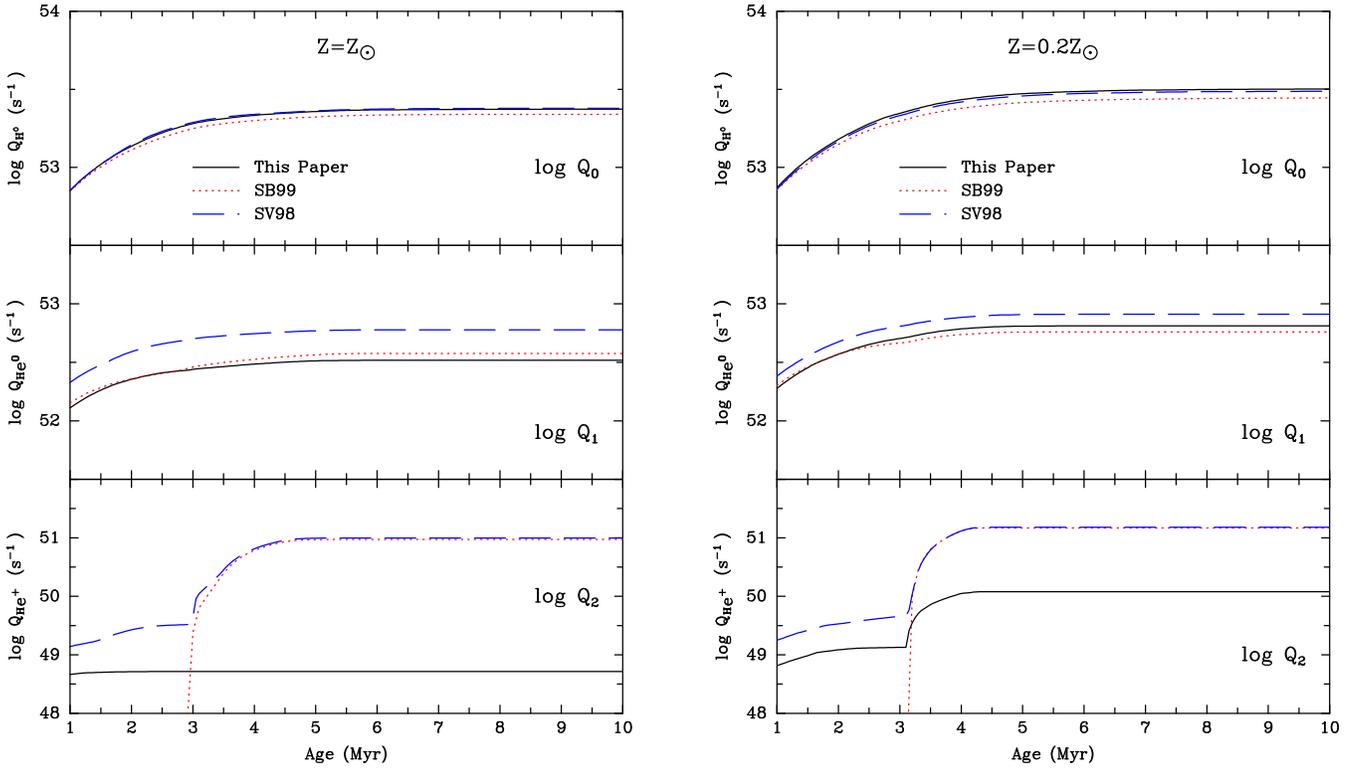

\begin{minipage}{8.4cm}
\psfig{file=fig_qsol_cns_col.eps,width=8.3cm,bbllx=77pt,bblly=207pt,bburx=437pt,bbury=643pt}
\end{minipage}
\hspace*{0.9cm}
\begin{minipage}{8.4cm}
\psfig{file=fig_qsmc_cns_col.eps,width=8.3cm,bbllx=77pt,bblly=207pt,bburx=437pt,bbury=643pt}
\end{minipage}
\caption{
The evolution of the photon luminosity for ages of 1--10\,Myr in the
ionizing continua of hydrogen ($\log Q_0$), He\one\ ($\log Q_1$) and He\two\
($\log Q_2$) at 0.2 and $Z_\odot$ for continuous star formation at time
steps of 0.1\,Myr. The details of the models are given in
Fig.~\ref{qs_inst}.}
\label{qs_cns}
\end{figure*}
\begin{figure*}
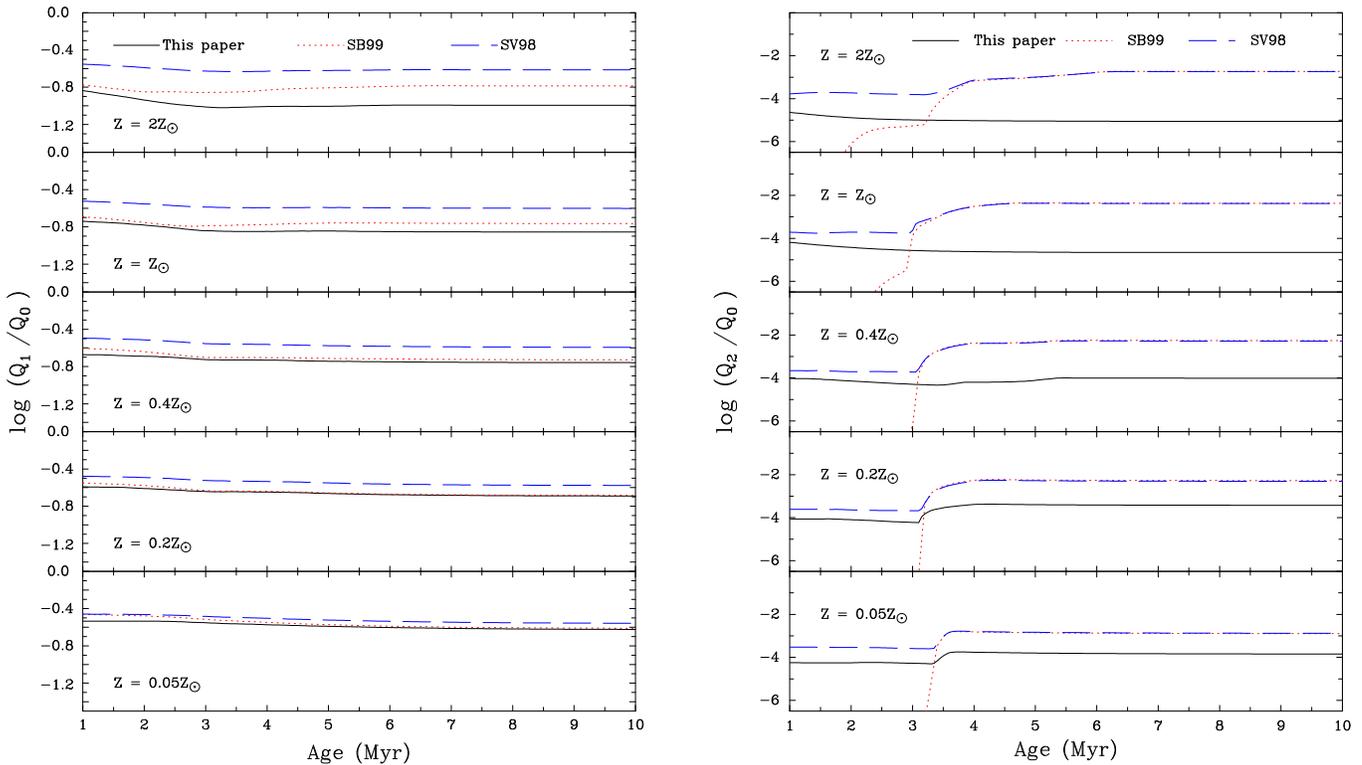

\begin{minipage}{8.4cm}
\psfig{file=fig_q1q0_cns_col.eps,width=8.3cm,bbllx=76pt,bblly=185pt,bburx=436pt,bbury=616pt}
\end{minipage}
\hspace*{0.9cm}
\begin{minipage}{8.4cm}
\psfig{file=fig_q2q0_cns_col.eps,width=8.3cm,bbllx=76pt,bblly=185pt,bburx=436pt,bbury=616pt}
\end{minipage}
\caption{The evolution of the hardness of the ionizing
spectra shown by the ratios of $\log (Q_1/Q_0)$ (lefthand side) and
$\log (Q_2/Q_0)$ (righthand side) at all five metallicities for
continuous star formation and ages of 1--10\,Myr. The details of the
models are given in Fig.~\ref{qs_inst}.}
\label{qrs_cns}
\end{figure*}

\subsection{Continuous star formation}\label{cns}
For continuous star formation, we assume a star formation rate of
1\,M$_\odot$\,yr$^{-1}$ and the same IMF parameters adopted for the
instantaneous case. In Fig.~\ref{sed_cns}, we show the spectral energy
distributions for continuous star formation.  In Figs.~\ref{qs_cns}
and \ref{qrs_cns}, the $\log Q$ values and their log ratios are shown.
After $\sim$ 5 \,Myr, a balance is achieved between massive star birth and
death and the ionizing fluxes become time independent.

The differences shown between the various models in Fig.~\ref{qs_cns}
emphasize the differences between non-LTE and LTE models and blanketed
vs. un-blanketed non-LTE models. Again, the predicted values of $Q_1$
are lower than SV98 because of our use of line-blanketed W-R models
and the lower ionizing fluxes of the WM-basic models. The difference
in $Q_2$ for ages $< 4\,$Myr between the SB99 and SV98 models is
simply due to LTE vs. non-LTE models being used for O stars.  The
increase in $Q_2$ when the first W-R stars are formed is much less
marked compared to SB99 and SV98 because of our use of line-blanketed
W-R models. The low value of $\log Q_2/Q_0 \approx -4$ implies that
nebular He\two\ will not be detectable at any metallicity from an
integrated stellar population undergoing continuous star formation
(Sect.~\ref{discuss}).

\section{Impact on Nebular Diagnostics}\label{photo}
We now assess the impact of the new grid of model atmospheres on the ionization
structure of H\two\ regions by considering single O star and synthetic
cluster models as a function of age. To perform these tests, we have used the
photoionization code {\sc cloudy} (version 96, Ferland 2002).

\subsection{Single star H\two\ regions}\label{htwo}
Many studies of the excitation of H\two\ regions have suggested that
the ionizing continuum becomes softer with increasing metal abundance
(e.g. Shields 1974; Stasi\' nska 1980; V\'\i lchez \& Pagel
1988). Shields \& Tinsley (1976) first suggested that this
relationship could be explained by a metallicity-dependent upper mass
limit for star formation. Recently, Bresolin et al. (1999) and
Kennicutt et al. (2000) have confirmed the decrease in effective
temperature with increasing metal abundance. They discuss the
possibility of a lower upper mass limit operating at abundances above
solar, although they caution that insufficient line blanketing in the
model atmospheres at high abundances may be the root cause. We are now
able to test this since our new O star grid contains fully
line-blanketed models up to 2\,Z$_\odot$.

The nebular line ratio He\one\ $\lambda5786$/H$\beta$ is a sensitive
diagnostic of the stellar temperature over the narrow range where
helium is partially ionized, corresponding to $\approx$
35\,000--40\,000\,K, but is insensitive to changes in the ionization
parameter $U$ (Bresolin et al. 1999).  Using the appropriate O dwarf
models listed in Table 1, we have calculated simple dust-free,
plane-parallel, ionization-bounded photoionization models with $\log
U=-3.0$ and a constant gas density $n=50$\,cm$^{-3}$ with a filling
factor of unity. We adopt the abundances and depletion factors given
by Dopita et al. (2000) for solar metallicity and scale these for
other metallicities. For the helium abundance as a function of
metallicity, we use the formula given by Dopita et al. (2000).  To
compare our results with the CoStar models, we have used the model
grid available within {\sc cloudy} corresponding to main sequence
stars for metallicities of 0.2\,Z$_\odot$ and $Z_\odot$, and
interpolated to match the temperatures of the WM-basic models. For
Kurucz LTE models, we have used models that are the closest match to
the WM-basic models in effective temperature and $\log g$.
\begin{figure}
\psfig{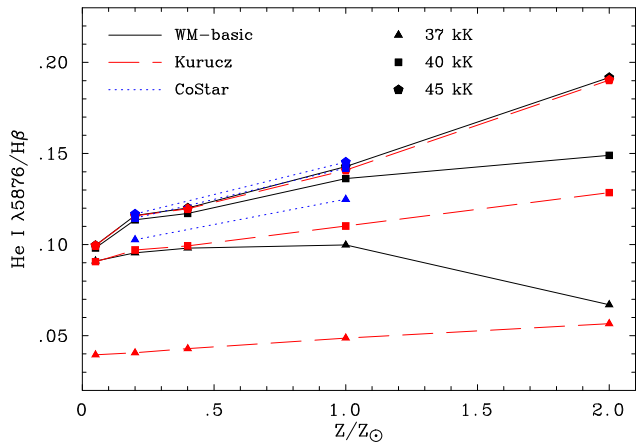}
\caption{The predicted nebular He\one\ $\lambda5786$/H$\beta$ ratio
plotted against metallicity for effective temperatures between 37--45kK
for WM-basic, CoStar and Kurucz O\,V model atmospheres.}
\label{hei}
\end{figure}

In Fig.~\ref{hei}, we plot He\one\ $\lambda5786$/H$\beta$ against
metallicity for three temperatures: 37\,000, 40\,000 and 45\,000\,K.
At the lower two temperatures, He is partially ionized and thus
sensitive to changing effective temperature. For the WM-basic models,
He becomes completely ionized at temperatures $\ge 43\,000$\,K.
Considering the lower two temperatures only, the Kurucz data points
increase gradually with abundance. Since the observed He\one\
$\lambda5786$/H$\beta$ ratio shows a decrease with increasing
abundance (e.g. Bresolin et al. 1999), this has led to the suggestion
that effective temperature decreases with increasing $Z$.  We see,
however, that the WM-basic model predictions are not flat but show a
decrease between 1 and 2\,Z$_\odot$. This is because our new
line-blanketed models have a softer ionizing flux in the He\one\
continuum at high $Z$ (Fig.~\ref{qrs_inst}) because of the deeper
line-core depths in non-LTE.  The strength of He\one\ $\lambda5876$
for a given effective temperature therefore decreases with increasing metal
abundance in a similar fashion to the observations.

The second significant difference between the WM-basic and the Kurucz
and CoStar models is the large offset in the effective temperature for
a given He\one\ line strength at a single metallicity. For example at
$Z_\odot$, at a temperature of 37\,000\,K, He\one\
$\lambda5786$/H$\beta =$ 0.10 (WM-basic); 0.12 (CoStar); and 0.05
(Kurucz). The absolute effective temperatures therefore depend heavily
on the model atmospheres employed.
\begin{figure*}
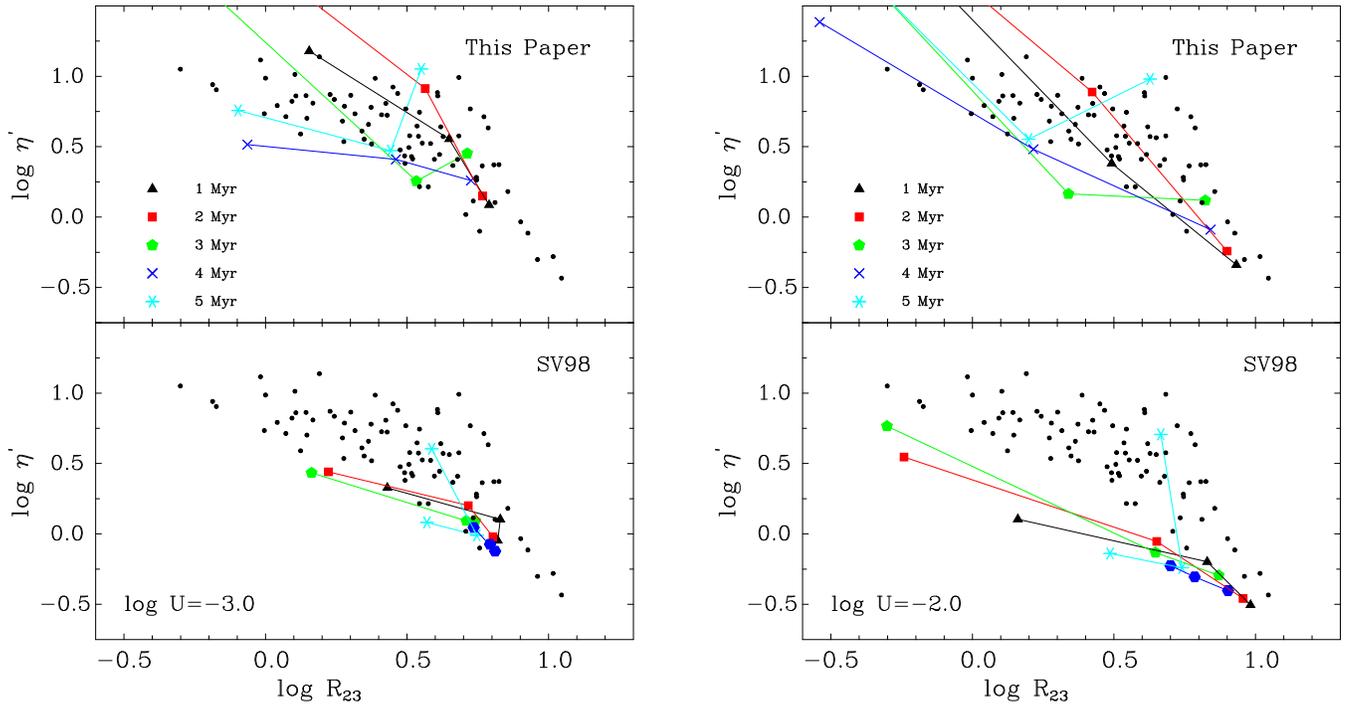

\begin{minipage}{8.4cm}
\psfig{file=fig_eta3_col.eps,width=8.3cm,bbllx=62pt,bblly=193pt,bburx=432pt,bbury=614pt}
\end{minipage}
\hspace*{0.9cm}
\begin{minipage}{8.4cm}
\psfig{file=fig_eta2_col.eps,width=8.3cm,bbllx=62pt,bblly=193pt,bburx=432pt,bbury=614pt}
\end{minipage}
\caption{The predicted strength of the softness parameter $\eta^\prime$ plotted
against the abundance indicator $R_{23}$ for 2, 1 and 0.2\,$Z_\odot$
for cluster models with $\log U=-3$ and $-2$ using (a) the models in this paper and (b) the
CoStar and SLG92 models of SV98.  The observed data points for the
extragalactic H\two\ region sample of Bresolin et al. (1999) are shown for
comparison.}
\label{eta}
\end{figure*}
\subsection{Cluster models}
Bresolin et al. (1999) compared observational diagnostic diagrams for
extragalactic H\two\ regions with the predictions of the cluster
models of Leitherer \& Heckman (1995). They found that the observed
line ratios could only be reproduced for ages of 1 and 2\,Myr, and
concluded that either the W-R ionizing continua are too hard, or the
H\two\ regions are disrupted at a young age.  More recently, Bresolin
\& Kennicutt (2002) studied a sample of metal-rich extragalactic
H\two\ regions. Some of these objects have W-R features in their
spectra but the nebular line ratios are not affected i.e.  they find no
observational evidence that H\two\ regions ionized by clusters
containing W-R stars show any signs of harder ionizing fluxes.

In Fig.~\ref{eta}, we plot the ``radiation softness'' parameter
$\eta^\prime$ defined by V\'\i lchez \& Pagel (1988) as
\[ \eta^\prime = \frac {[{\rm O\,II}]\,\lambda\lambda3726,3729/
{\rm [O\,III]}\,\lambda\lambda4959,5007}
{{\rm [S\,II]}\,\lambda\lambda6717,6731/
{\rm [S\,III]}\,\lambda\lambda9069,9532}\]
against the abundance parameter $R_{23}=$([O II] $\lambda3727 + $[O
III] $\lambda\lambda4959,5007)/$H$\beta$. The $\eta^\prime$ parameter
is a sensitive measure of the softness of the radiation field and
provides an excellent test of the correctness of our new models.  We
show the predicted values of $\eta^\prime$ for metallicities of 2, 1
and 0.2\,Z$_\odot$ and ages of 1--5\,Myr (i.e. up to the end of the
W-R phase) for an instantaneous burst model of mass $10^6$\,M$_\odot$.
We have used the same photoionization model parameters as in
Sect.~\ref{htwo} and since $\eta^\prime$ is sensitive to the
ionization parameter, we show two cases with $\log U=-3$ and $-2$. We
compare the predictions of the new grid with the CoStar and SLG92 W-R
models of SV98 and the observational data of Bresolin et
al. (1999). It can be seen that the SV98 models are too hard,
particularly for the $\log U=-2.0$ case, whereas the new models are
coincident with the data points.  We conclude that, for the parameter
range explored, the ionizing fluxes of the new models are in much
better agreement with the observed emission line ratios of H\two\
regions than previous model grids.

\section{Discussion}\label{discuss}
The success of the technique of evolutionary synthesis for deriving
the properties of young stellar populations depends on the accuracy of
the model atmospheres and evolutionary tracks for massive stars.  In
this paper, we have sought to provide a grid of realistic ionizing
fluxes for O and W-R stars using the latest line-blanketed, non-LTE
atmosphere codes. We have carefully chosen the input stellar
parameters to reflect current values available in the literature.
While these appear to be well established for the O spectral class,
there have been major revisions for the W-R stars. We have
incorporated these changes by using lower clumping-corrected mass loss
rates and lower temperatures compared to the previous W-R grid of
SLG92.  We have also introduced metallicity-dependent mass loss rates
because recent observational and theoretical work indicate that the
strength of W-R winds should scale with heavy metal abundance
(Crowther et al. 2002a).

The lower, metallicity-dependent, W-R mass loss rates have implications
for the amount of mass and energy returned to the interstellar medium
from young starbursts. We will address this topic in a future paper.
Briefly, in comparison to Leitherer et al. (1992), we find that
W-R stars no longer dominate the energy input, particularly at
low metallicity. Our revised O star mass loss rates, however, are
larger than those given by the analytical formula of Leitherer
et al. (1992), and this leads to an overall increase in the rates
of mass and energy input.

As discussed in Section~\ref{comp} and SLG92, the emergent flux below
228\,\AA\ depends on the wind density and hence the scaling law used
to scale the wind parameters with metallicity.  The precise scaling
even for O stars has not been determined empirically and relies on the
predictions of radiatively-driven wind theory. It is of course unknown
for W-R stars and we simply chose to use the O star scaling law. The
predicted number of He$^+$ ionizing photons at metallicities below
solar should therefore be considered as rather uncertain.

We now discuss the presence of nebular He\two\ $\lambda4686$ emission.
Guseva et al. (2000) compile a list of 30 H\two\ galaxies
with nebular He\two\ detections. Schaerer (1996) used CoStar and the
SLG92 models to synthesize the strength of nebular He\two\
$\lambda4686$ emission in the early stages of a starburst. Schaerer
predicts that strong nebular He\two\ emission will be present during
the W-R phase with $I$(He\two)/$I$(H$\beta$)$=0.01$--0.025 for
$0.2$\,Z$_\odot \le Z \le Z_\odot$. Guseva et al. (2000) find that the
predicted He\two\ emission line strengths (as given by SV98) agree
reasonably well with the observations except that the SV98 models
predict the highest $I$(He\two)/$I$(H$\beta$) ratios to occur at 1 and
2\,$Z_\odot$, while no nebular He\two\ is detected in objects with
$Z>0.2$\,Z$_\odot$. They also find that W-R stars cannot be the sole
source of nebular He\two\ emission since only 60 per cent of their
sample have W-R stars present, and they suggest additional mechanisms
such as radiative shocks are needed to explain all the observations.

Assuming case B recombination and an ionization-bounded nebula, the
strength of He\two\ $\lambda4686$ relative to H$\beta$ is given by
$I$(He\two)/$I$(H$\beta$)$=2.14\,Q_2/Q_0$ (SV98). If we further assume
that the minimum detectable $I$(He\two)/$I$(H$\beta$)$=0.01$, then
$\log Q_2/Q_0 \ge -2.33$ for He\two\ $\lambda4686$ to be
observable. From Fig.~\ref{qrs_inst}, the highest value of $\log
Q_2/Q_0 = -2.4$, corresponding to $I$(He\two)/$I$(H$\beta$)$=0.08$,
occurs at the beginning of the W-R phase at 0.2\,Z$_\odot$. Thus any
nebular He\two\ will be marginally observable.  With the current grid
of line-blanketed W-R models, it seems unlikely that the observed
nebular He\two\ associated with W-R galaxies is in general due to W-R
star photoionization.

We must emphasize that the W-R grid is based on the observational
parameters of ``normal or average'' W-R stars and thus excludes the
rare, hot early WN and WO stars with weak winds and significant fluxes
above 54\,eV. Since we do not understand the reasons for their extreme
parameters, it is impossible to predict their impact on the ionizing
output of young starbursts. Given their rarity in the Galaxy and LMC,
we expect their influence to be negligible. But, if for example, they
are formed via the binary channel in significant numbers at low $Z$,
they may become an important source of He$^+$ ionizing photons.  We
have also neglected shocks in the stellar winds; they will increase
the ionizing flux in the EUV.  It is not known how the strength of
these shocks (caused by instabilities in the line-driving force) will
scale with metallicity.

The problem of assigning W-R atmospheres to evolutionary models is a
long-standing one and has been extensively discussed by SLG92 and in
Section~\ref{W-Rcomp}. To connect our new W-R models to the
evolutionary tracks, we matched $T_*$ to $0.6 T_{\rm {hyd}} + 0.4
T_{2/3}$ rather than matching $T_*$ directly to $T_{\rm {hyd}}$, as in
Leitherer et al. (1999) because of the lower temperatures of the new
W-R grid. The differences we find for starbursts in the W-R phase
compared to the results of SV98 and Leitherer et al. (1999) is
therefore potentially a combination of the effects of line blanketing
and using lower temperature W-R models. The overall effect of the
method we have used to connect the new W-R grid with the evolutionary
tracks can be gauged from Figs.~\ref{qs_inst} and \ref{qrs_inst}. The
relative heights of the WN and WC bumps at $\sim$ 3.5 and 5\,Myr are
very similar for all three comparisons. This suggests that the main
differences we find in the W-R phase are primarily due to the
inclusion of line-blanketing and that the lower temperatures are a
secondary effect. The formalism that we used to interface the W-R grid
to the evolutionary tracks should, however, be considered as rather
uncertain.  For this reason, it is treated as a free parameter in the
up-dated Starburst99 code. 

Finally, we mention that the predictions we have made for the W-R phase
as a function of metallicity depend on the evolutionary tracks of
Meynet et al. (1994) for single stars with enhanced mass loss
rates. Leitherer (1999) has compared various W-R synthesis models and
discusses the differences in single and binary models. He finds that
the W-R/O and WC/WN ratios are highest when the enhanced mass loss
tracks are used, and that the W-R phase extends to between 6 and
10\,Myr when binary formation is included (see also SV98). Recent
evolutionary models including rotation (Maeder \& Meynet 2000) have
removed the necessity of artificially increasing the mass loss rate to
obtain better agreement with observations of W-R stars. The lifetime of
the W-R phase is increased, the WC/WN ratio is reduced, and more W-R
stars are produced at low $Z$ (Maeder \& Meynet 2001). The next stage
in improving evolutionary synthesis models should be the inclusion of
evolutionary models with rotation.

\section{Conclusions}\label{conc}
We have presented a large grid of non-LTE, line-blanketed models for O
and W-R stars covering metallicities of 0.05, 0.2, 0.4, 1 and
2\,Z$_\odot$. The grid is designed to be used with the evolutionary
synthesis code Starburst99 (Leitherer et al. 1999) and in the analysis
of H\two\ regions ionized by single stars. We have computed 110 models
for O and early B stars using the WM-basic code of Pauldrach et
al. (2001).  The OB stellar parameters are defined at solar
metallicity and are based on the most recent compilations of
observational data.  The mass loss rates and terminal velocities have
been scaled according to metallicity by adopting power law exponents
of 0.8 and 0.13 (Leitherer et al. 1992).

For the W-R grid, we used the model atmosphere code {\sc cmfgen} of
Hillier \& Miller (1998) and computed 60 WN and 60 WC models. The new
grid is based on what we consider to be the most realistic parameters
derived from recent observations and individual model atmosphere
analyses. In particular, the upper temperature limit of the grid is
140\,000\,K since the vast majority of W-R stars that have been
analysed fall below this value. For the W-R mass loss rates at
solar metallicity, we used the relationships derived by Nugis \&
Lamers (2000) which are corrected for inhomogeneities in the W-R winds.
For the first time, we have introduced a W-R wind/metallicity
dependence and adopted the same power law exponents used for the O
star models.  We argue in Sect.~\ref{W-Rgrid} that recent theoretical
and observational work indicate that the strengths of W-R winds must
depend on metallicity. We stress that the new W-R grid excludes the few
known examples of individual hot W-R stars with weak winds that have
significant ionizing fluxes above 54\,eV. Since they are so rare (in
the Galaxy and LMC, at least), we expect them to make a negligible 
contribution to the ionizing flux of a young starburst.

We find significant differences in the emergent fluxes from the
WM-basic models compared to the CoStar models of Schaerer \& de Koter
(1997). For supergiants, the wind density determines the transparency
below 228\,\AA.  Generally, we find a lower flux in the He\one\ 
continuum with important implications for nebular line diagnostic
ratios.  We believe that the CoStar models over-predict the number of
He$^0$ ionizing photons through the neglect of photon absorption in
line transitions and their re-emission at longer wavelengths (Crowther
et al.  1999).

We compared the new W-R model emergent fluxes with the pure helium W-R
models of Schmutz et al. (1992) which have very different parameters
(particularly wind densities and temperatures). We find that at
$\sim 45\,000$\,K for solar metallicity WN models, the
emergent flux below 504\,\AA\ is much lower than the SLG92 models
because of the inclusion of line blanketing. At 60\,000\,K, we find
that blanketing is less important, and at 90\,000\,K, the wind density
controls the emergent flux below 228\,\AA.

The WM-basic model grid has been integrated into Starburst99
(Leitherer et al. 1999) by directly replacing the Lejeune et
al. (1997) LTE model library. For the W-R grid, we have used a new
method of matching the models to the evolutionary tracks by taking a
weighted mean of the uncorrected hydrostatic temperature $T_{\rm
{hyd}}$ and the corrected hydrostatic temperature $T_{2/3}$. This is
necessary because of our lower, more realistic temperatures,
particularly for the WC stars.

We next compared the output ionizing fluxes of the new grid integrated
into Starburst99 with the evolutionary synthesis models of Leitherer
et al. (1999) and Schaerer \& Vacca (1998) for an instantaneous burst
and continuous star formation. The changes in the ionizing outputs are
dramatic, particularly during the W-R phase, with the details depending on
metallicity because of line-blanketing and wind density effects.
For an instantaneous burst, we find that the number of He$^+$ ionizing
photons ($Q_2$) emitted during the W-R phase is negligible at $Z_\odot$
and higher. At lower metallicities, $Q_2/Q_0$ is softer by a factor of
$\sim 20$ compared to the SLG92 models. In contrast to Schaerer
(1996), we predict that nebular He\two\ will be at, or just below, the
detection limit in low metallicity starbursts during the W-R phase.  We
also find lower He$^0$ ionizing fluxes for $Z\ge 0.4$\, Z$_\odot$ and
ages of $\le 7$\,Myr compared to SV98 because of the smaller
contributions of the line-blanketed models. The ionizing fluxes of the
continuous star formation models emphasize the differences found for
the single star models.

We have tested the correctness of the new model grid by computing
nebular line diagnostic ratios using {\sc cloudy} and single star and
evolutionary synthesis models as inputs. For the former, we calculated
the nebular He\one\ $\lambda 5786$/H$\beta$ ratio as a function of $Z$
since this is a sensitive diagnostic of $Q_1/Q_0$ for the temperature
range where He is partially ionized. Observations indicate that this
ratio decreases with increasing metal abundance, leading to the
suggestion that the effective temperature decreases with increasing
$Z$, and thus that the upper mass limit may be $Z$-dependent
(e.g. Bresolin et al. 1999). The WM-basic models for dwarf O stars
show a decrease in He\one\ $\lambda 5786$/H$\beta$ above $Z_\odot$ in
contrast to Kurucz LTE models which are essentially independent of
$Z$. This decrease is in the same sense as the observations and
suggests that the observed decline in He\one\ $\lambda 5786$/H$\beta$
with increasing $Z$ is simply due to the effect of line blanketing
above $Z_\odot$, and is not caused by a lowering of the upper mass
limit.

The ionizing fluxes of the instantaneous burst models have been tested
by comparing them to the predictions of SV98 and the observations of
Bresolin et al. (1999) for extragalactic H\two\ regions.  In plots of
the softness parameter $\eta^\prime$ against the abundance indicator
$R_{23}$ for ages of 1--5\,Myr and $\log U=-2$ and $-3$, we find that
the new models cover the same parameter space as the data points in
contrast to the SV98 ionizing fluxes which are generally too hard at
all ages.

We therefore conclude that the ionizing fluxes of the new model grid
of O and W-R stars represent a considerable improvement over the model
atmospheres that are currently available in the literature for massive
stars. To prove their worth, they should be tested rigorously against
specific observations of single H\two\ regions and young bursts of
star formation.

The grid of ionizing fluxes for O and W-R stars and the updated
Starburst99 code can be obtained from: 
{\tt http://www.star.ucl.ac.uk/starburst}.

\section*{Acknowledgements}
We especially thank Adi Pauldrach and John Hillier for the use of
their model atmosphere codes. We also thank Fabio Bresolin, Claus
Leitherer and Daniel Schaerer for many useful comments which
significantly improved this paper. PAC acknowledges financial support
from the Royal Society.

\bsp
\label{lastpage}
\end{document}